\documentclass[sigconf,9pt]{acmart}
\pdfoutput=1

\usepackage{tikz}
\usepackage{amsmath}
\usepackage{pifont}
\usepackage{subcaption}
\usepackage{multirow}
\usepackage{algorithm}
\usepackage{algorithmic}
\usepackage[normalem]{ulem}
\usepackage{nicefrac}
\usepackage{graphicx}
\usepackage{todonotes}

\usepackage{xcolor}

\newcommand{\argmin}{\arg\!\min}

\newcommand{\sect}{\textsection}

\AtBeginDocument{%
  }

\acmYear{2024}\copyrightyear{2024}
\setcopyright{rightsretained}
\acmConference[MOBISYS '24]{The 22nd Annual International Conference on Mobile Systems, Applications and Services}{June 3--7, 2024}{Minato-ku, Tokyo, Japan}
\acmBooktitle{The 22nd Annual International Conference on Mobile Systems, Applications and Services (MOBISYS '24), June 3--7, 2024, Minato-ku, Tokyo, Japan}
\acmDOI{10.1145/3643832.3661854}
\acmISBN{979-8-4007-0581-6/24/06}

\begin{document}

\title{Invisible Optical Adversarial Stripes on Traffic Sign against Autonomous Vehicles}

\author{Dongfang Guo}
\affiliation{%
  \institution{Nanyang Technological University}
  \country{Singapore}
}
\email{dongfang.guo@ntu.edu.sg}

\author{Yuting Wu}
\affiliation{%
  \institution{Nanyang Technological University}
  \country{Singapore}
}
\email{yuting.wu@ntu.edu.sg}

\author{Yimin Dai}
\affiliation{%
  \institution{Nanyang Technological University}
  \country{Singapore}
}
\email{yimin006@e.ntu.edu.sg}

\author{Pengfei Zhou}
\affiliation{%
  \institution{University of Pittsburgh}
  \city{Pittsburgh}
  \country{USA}
}
\email{pengfeizhou@pitt.edu}

\author{Xin Lou}
\affiliation{%
  \institution{Singapore Institute of Technology}
  \country{Singapore}
}
\email{lou.xin@singaporetech.edu.sg}

\author{Rui Tan}
\affiliation{%
  \institution{Nanyang Technological University}
  \country{Singapore}
}
\email{tanrui@ntu.edu.sg}
\renewcommand{\shortauthors}{Dongfang Guo, Yuting Wu, Yimin Dai, Pengfei Zhou, Xin Lou, and Rui Tan.}

\begin{abstract}
  Camera-based computer vision is essential to autonomous vehicle's perception. This paper presents an attack that uses light-emitting diodes and exploits the camera's rolling shutter effect to create adversarial stripes in the captured images to mislead traffic sign recognition. The attack is stealthy because the stripes on the traffic sign are invisible to human. For the attack to be threatening, the recognition results need to be stable over consecutive image frames. To achieve this, we design and implement \emph{GhostStripe}, an attack system that controls the timing of the modulated light emission to adapt to camera operations and victim vehicle movements. Evaluated on real testbeds, GhostStripe can stably spoof the traffic sign recognition results for up to 94\% of frames to a wrong class when the victim vehicle passes the road section. In reality, such attack effect may fool victim vehicles into life-threatening incidents. We discuss the countermeasures at the levels of camera sensor, perception model, and autonomous driving system.
\end{abstract}



\maketitle

\section{Introduction}
\label{sec:intro}

Camera-based computer vision is an essential perception channel of autonomous vehicles, especially for the tasks of traffic sign recognition and lane detection \cite{marti2019review}.
Thus, reliable camera-based perception is vital to autonomous vehicle's safety. 
Recent research on adversarial examples \cite{goodfellow2014explaining,adversarial-example-list} has aroused the consciousness regarding the potential vulnerability of camera-based perception. 
To better understand its security in the context of autonomous driving, this paper presents a physically deployable and stealthy optical adversarial-example attack that exploits the camera's rolling shutter effect to fool the car's traffic sign recognition.

Camera sensors are based on either charge coupled device (CCD) or complementary metal oxide semiconductor (CMOS). CCD sensor captures the entire frame by exposing all pixels simultaneously. Differently, CMOS sensor captures the image in a line-by-line manner using an electronic rolling shutter. Thus, the lines of a frame are exposed during different time periods. Compared with CCD, CMOS is less costly.
As CMOS provides a satisfactory balance between cost and image quality, it has been widely adopted in camera products, including those deployed on vehicles. For instance, both Tesla and Baidu Apollo use CMOS cameras in their designed vehicles \cite{teslahardware,apollohardware}.

Despite its advantages, CMOS camera exhibits {\em rolling shutter effect} (RSE) \cite{rollingshutterproblem} when the input light contains flickering frequencies close to the operational frequency of the rolling shutter.
Specifically, as the rows of a CMOS sensor are exposed in slightly different time periods, rapid changes of the input light can introduce varied color shades in different sensor scanlines and thus image distortion. Recent studies have shown the security implication of RSE, i.e.,
attackers can control or perturb the input light to create colored stripes on the captured image to mislead the computer vision's interpretation of the image.
A recent work \cite{sayles2021invisible} uses light-emitting diodes (LEDs) to create flickering ambient illumination and mislead the classification of the images taken in the space under attack. In \cite{kohler2021they}, a laser beamed into camera lens creates colored stripes to disrupt object detection.

\begin{figure}[t!]
  \centering
  \begin{subfigure}[t]{\linewidth}
    \includegraphics[width=\linewidth]{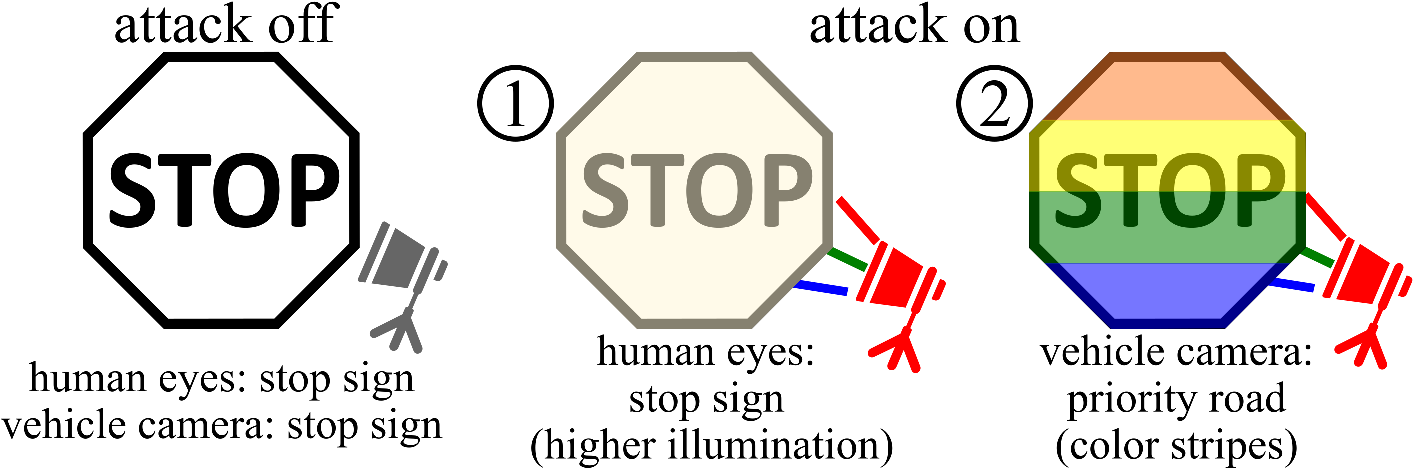}
    \vspace{-1.5em}
    \caption{Invisible perturbation.}
    \label{sfig:intro_invisible}
  \end{subfigure}
  \hfill
  \begin{subfigure}[t]{\linewidth}
  \centering
  \vspace{0.5em}
  \includegraphics[width=\linewidth]{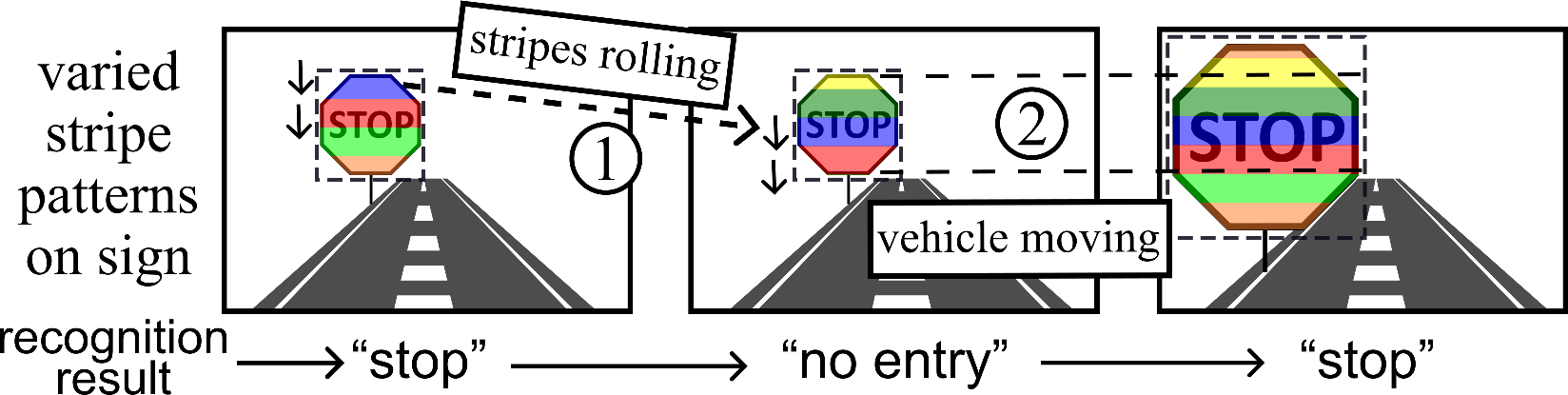}
  \vspace{-1.9em}
    \caption{Unstable attack.}
    \label{sfig:intro_unstable}
  \end{subfigure}
  \hfill
  \begin{subfigure}[t]{\linewidth}
    \centering
    \vspace{0.8em}
    \includegraphics[width=\linewidth]{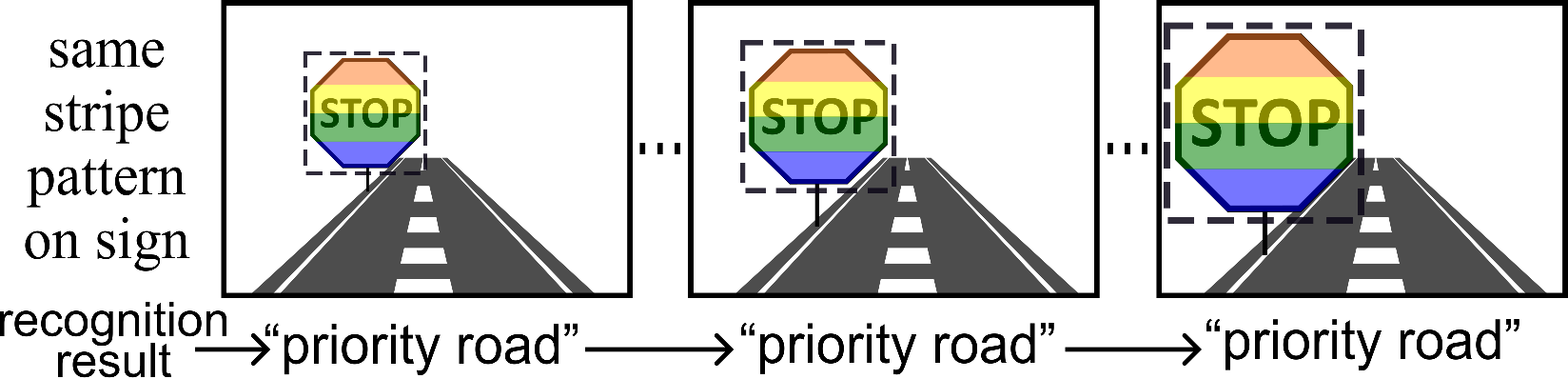}
    \vspace{-1.9em}
      \caption{Stable attack.}
      \label{sfig:intro_stable}
  \end{subfigure}
  \vspace{-1em}
  \caption{Invisible optical adversarial-example attack against traffic sign recognition.}
  \label{fig:intro_fig}
\end{figure}

While the existing studies have implemented elementary RSE attacks on single image frames captured in controlled environments, they fall short of achieving stable attack results over a sequence of frames. This paper aims to achieve stable attack results which render clearer security implications in the autonomous driving context.
In the envisaged attack as illustrated in Fig.~\ref{sfig:intro_invisible}, an LED is deployed in the proximity of a traffic sign plate and projects controlled flickering light onto the plate surface. As the flickering frequency is beyond human eye's perception limit (up to $50$-$90\,\text{Hz}$ \cite{mankowska2021critical}), the flickering is invisible to human and the LED appears as a benign illumination device, as illustrated in Fig.~\ref{sfig:intro_invisible}-\textcircled{\small{1}}. 
Meanwhile, on the image captured by the camera, as illustrated in Fig.~\ref{sfig:intro_invisible}-\textcircled{\small{2}}, the RSE-induced colored stripes mislead the traffic sign recognition.
For the attack to mislead the autonomous driving program to make erroneous decisions unconsciously, the traffic sign recognition results should be wrong and same across a sufficient number of consecutive frames. 
We call the attack meeting this requirement {\em stable}. 
If the attack is not stable, an anomaly detector may identify the malfunction of the recognition and activate a fail-safe mechanism, e.g., falling back to manual driving or emergency safe stopping, rendering the attack less threatening.

Implementing a stable attack is 
a non-trivial task that necessitates addressing two essential challenges,
as illustrated in Figs.~\ref{sfig:intro_unstable} and \ref{sfig:intro_stable}. First, the stable attack requires the capability of stablizing the appearance of the pre-designed colored stripes on the image cropout containing the traffic sign.
Otherwise, if the stripes captured by the camera roll on the traffic sign (e.g., rolling downwards in Fig.~\ref{sfig:intro_unstable}-\textcircled{\small{1}}), the recognition result will change over time.
The rolling is caused by the discrepancy between the LED flickering frequency and the camera's rolling shutter frequency.
Thus, the stripe position stabilization requires precise calibration of LED's flickering frequency.
Second, the stable attack must adapt to the time-varying position and size of the traffic sign cropout within the original image sequence captured by the moving victim vehicle.
Otherwise, the stripe pattern on the traffic sign will change over time. For instance, in Fig.~\ref{sfig:intro_unstable}-\textcircled{\small{2}}, when the stripes keep still in the field of view (FoV), the varying sign in the FoV contains varying stripe patterns, leading to varying recognition results.
Thus, a stable attack, as illustrated in Fig.~\ref{sfig:intro_stable}, needs to carefully control the LED's flickering based on the information about the victim camera's operations and real-time estimation of the traffic sign position and size in the camera's FoV.

To address the aforementioned challenges in crafting a stable attack, this paper presents the designs of two versions of an attack system called \emph{\textbf{GhostStripe}}
with different requirements on the attack deployment.
The first version, \emph{\textbf{GhostStripe1}}, maintains stationary adversarial stripes in the FoV by calibrating the LED flickering frequency.  
GhostStripe1 employs a {\em vehicle tracker} to monitor the victim vehicle's real-time location and dynamically adjusts the LED flickering accordingly. 
GhostStripe1 does not require any instrumentation on the victim vehicle. It aims to maintain the victim's traffic sign recognition result stable over time. 
However, it is an untargeted attack, in that the recognition result is unpredictable because the vertical positions of the adversarial stripes  are not controlled by the attacker.
To achieve targeted attack (i.e., the attacker can control the victim's recognition result), on top of GhostStripe1, \emph{\textbf{GhostStripe2}} deploys a {\em framing sniffer} to sense the victim camera's framing moments via a current transducer clipped on the power wire of the camera. The sniffer transmits the detected framing moments to the LED controller to refine the timing control of the flickering.
Although installing the framing sniffer requires physical access to the victim vehicle, it is possible, say, during maintenance by an auto care provider colluding with the attacker.

The main contributions of this paper are as follows:
\begin{itemize}
\item We analyze the principles for achieving stable RSE-based optical adversarial-example attack against autonomous driving perception and present techniques to satisfy the conditions obtained from the analysis. 
\item Following the principles, we design GhostStripe, a physically deployable attack system. Two versions of GhostStripe are designed to enable untargeted and targeted attacks with different attack deployment requirements, respectively. 
\item We evaluate GhostStripe on a real outdoor testbed and a lab testbed with Leopard Imaging AR023ZWDR as the victim camera, which is used in Baidu Apollo's hardware reference design \cite{apollohardware}. 
On the outdoor testbed, GhostStripe1 and GhostStripe2 can achieve up to 94\% and 97\% success rates in launching untargeted and targeted attacks, respectively.
\end{itemize}

{\em Paper organization:}
\sect\ref{sec:bg} introduces background and preliminaries.
\sect\ref{sec:prob_state} and \sect\ref{sec:system_design} design and implement GhostStripe, respectively.
\sect\ref{sec:testbed_and_prototype} describes the testbeds.
\sect\ref{sec:evaluation} presents experiment results.
\sect\ref{sec:countermeasures} discusses possible countermeasures.
\sect\ref{sec:discussion} discusses several issues.
\sect\ref{sec:related_work} reviews related work.
\sect\ref{sec:conclusion} concludes this paper.

\section{Background and Preliminaries}
\label{sec:bg}

\subsection{Traffic Sign Recognition}
\label{subsec:bg_recognition_in_av}

Car-borne camera-based traffic sign recognition consists of detection and classification phases \cite{zhu2016traffic,yang2015towards}, which are usually based on deep neural networks (DNNs). First, the detector locates the traffic sign in the image frames.
Then, the detected traffic signs are cropped and fed to the classifier for interpretation. In this paper, we focus on compromising the classifier. Evaluation in \sect\ref{subsubsec:impact_on_detection} shows that GhostStripe has negligible impact on the detector.

\subsection{Rolling Shutter Operation and Effect}
\label{subsec:rs_mechanism}

\begin{figure}
    \centering
    \includegraphics[width=\linewidth]{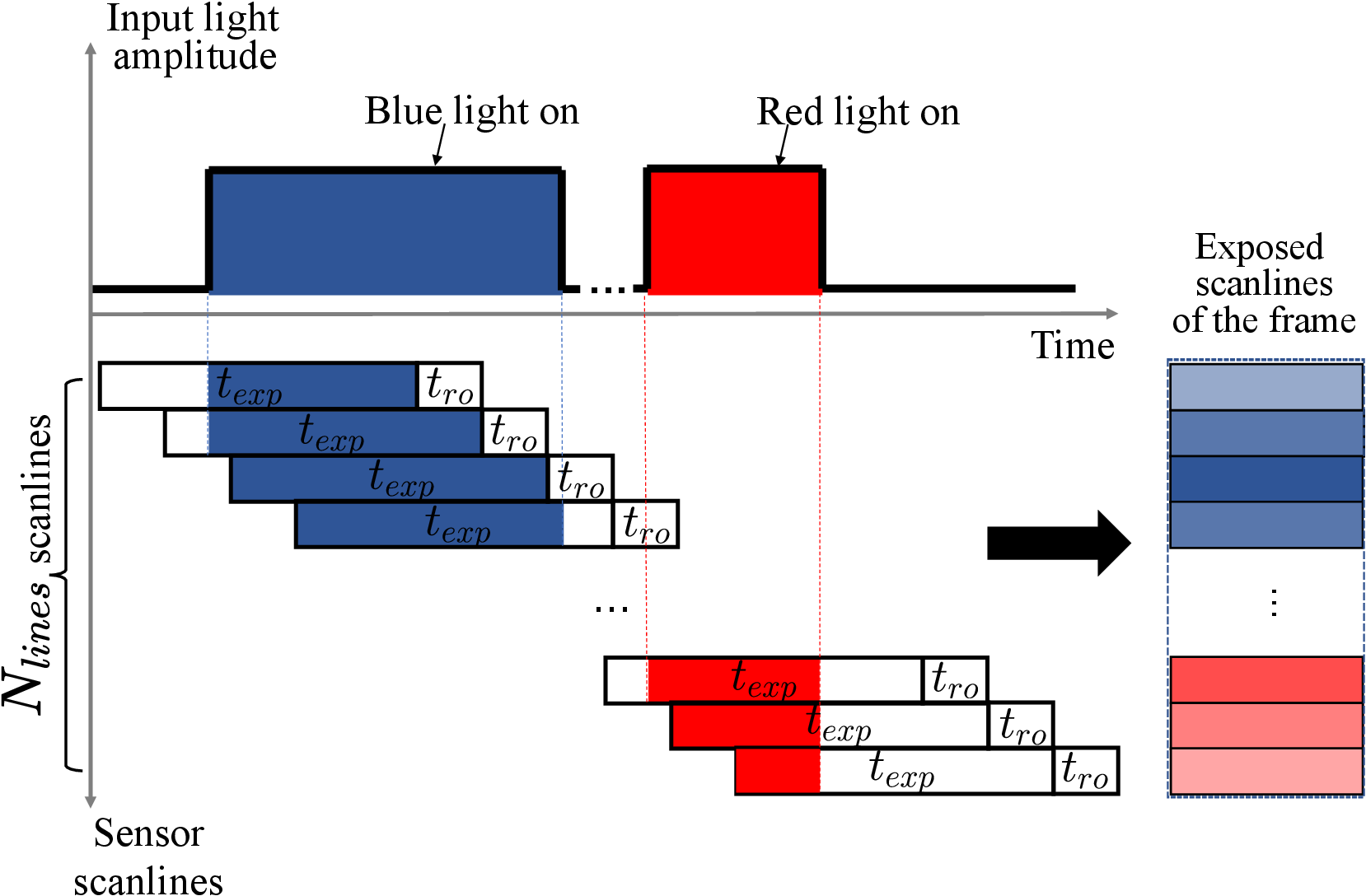}
    \vspace{-2em}
    \caption{Rolling shutter's operation and RSE.}
    \label{fig:rs_mechanism}
\end{figure}

Fig.~\ref{fig:rs_mechanism} illustrates the rolling shutter's operation. As CMOS sensor typically has no memory buffer to store the charge in the photodiode array, it exposes and reads out the pixel values on a row-wise basis, typically from top to bottom. Denote by $N_{lines}$ the number of scanlines. 
When capturing an image frame, each scanline is exposed for a time period $t_{exp}$. After that, the data of the scanline is read out within a readout time denoted by $t_{ro}$. As illustrated in Fig.~\ref{fig:rs_mechanism}, the exposure-readout processes for the scanlines are pipelined. The process for the next scanline is $t_{ro}$ in time later than that of the previous scanline.
As a result, the total time for capturing a frame is
$t_{cap}= N_{lines} \times t_{ro} + t_{exp}$.
Note that $t_{ro}$ is fixed and can be found from the sensor specification.
The $t_{exp}$ is fixed for a certain frame but can vary across frames depending on the camera's exposure setting. The following terms are defined for the rest of this paper. {\em Framing moment} is the time instant at which the exposure of the first scanline starts. {\em Frame period} denoted by $T_{frame}$ is the time between the framing moments of two consecutive frames, which is the reciprocal of the camera's frame rate. We have $T_{frame} \ge t_{cap}$.

Now, we explain the formation of RSE. As shown in Fig.~\ref{fig:rs_mechanism}, two light pulses (a blue pulse and a red pulse) affect the captured image. A pulse affects the scanlines exposed during the pulse time. The intensity of the affection on a scanline
depends on the amount of the pulse time within the scanline's exposure time. Consequently, the light pulses result in horizontal stripes in the captured frame.

\subsection{RSE-Based Adversarial Examples}
\label{subsec:cvpr_paper}

An adversarial example, which is the sum of the original sample and a minute perturbation, misleads a DNN to produce a result different from that of the original sample \cite{goodfellow2014explaining}.
The work \cite{sayles2021invisible} presents a method that controls the LED flickering to create RSE-induced stripes as the adversarial perturbation to mislead an object recognition DNN. Its essence is as follows.
Denote by $c \in \{R, G, B\}$ the color channel. We use $c$ as the superscript of the quantity defined for a certain color channel. Denote by $t \in [0, t_{cap}]$ the relative time starting from the current frame's framing moment, by $f^c(t) \in [0, 1]$ the LED's relative emission intensity, by $\alpha^c$ the ambient light intensity, by $\beta^c$ the LED's maximum intensity, by $l_{tex}^c(u,v)$ the texture of the scene, where $(u,v)$ are the coordinates in the camera's FoV. 
Illuminated by both the ambient light and LED, the light intensity in color channel $c$ at position $(u,v)$ in the scene at time $t$ is $l_{tex}^c(u,v) \cdot (\alpha^c + \beta^c f^c(t))$. From Fig.~\ref{fig:rs_mechanism}, the exposure of the $v$th scanline starts at time instant $vt_{ro}$. Thus, the value of pixel $(u,v)$ in color channel $c$ is given by
\vspace{-1.0em}
\begin{align*}
    I^c(u,v) \!&=\! \rho \int_{vt_{ro}}^{vt_{ro} + t_{exp}}\!\!\! l_{tex}^c(u,v)(\alpha^c + \beta^c f^c(t)) \mathrm{d}t \\
    &= I_{amb}^c(u,v) + I_{att}^c(u,v) g^c(v)
\end{align*}
where $\rho$ is the sensor gain, $I_{amb}^c(u,v) = \rho l_{tex}^c(u,v) t_{exp} \alpha^c$, $I_{att}^c(u,v) = \rho l_{tex}^c(u,v) t_{exp} \beta^c $, $g^c(v)=\frac{1}{t_{exp}} \int_{vt_{ro}}^{vt_{ro} + t_{exp}} f^c(t) \mathrm{d}t$. Note that $I_{amb}^c(u,v)$ is the image in color channel $c$ captured with ambient illumination only. The $I_{att}^c(u,v)$ is the image captured with light emitted from the LED in full intensity all the time and no ambient illumination.
It can be obtained by $I_{att}^c(u,v)=I_{full}^c(u,v)-I_{amb}^c(u,v)$, where $I_{full}^c(u,v)$ is the image captured with both the ambient illumination and the full-intensity light from the LED. Both $I_{amb}^c(u,v)$ and $I_{full}^c(u,v)$ are collected by the attacker in advance.
The LED control signal in all color channels $f(t)=\{f^R(t), f^G(t), f^B(t)\}$ is designed by solving 
$\argmin_{f(t)} \displaystyle \ell(\mathcal{M}(I(u,v)), k)$,
where $I(u,v)=\{I^R(u,v), I^G(u,v), I^B(u,v)\}$, $\mathcal{M}(\cdot)$ is the classifier, $k$ is the target class of the attack (i.e., the attack aims to mislead the classifier to produce class $k$), $\ell(\mathcal{M}(I(u,v)), k)$ is the classification loss for the target class $k$ when the classifier is fed with $I(u,v)$.

\section{Design Principles of GhostStripe}
\label{sec:prob_state}

\begin{figure*}[ht!]
  \vspace{-1em}
  \centering
  \includegraphics[width=\linewidth]{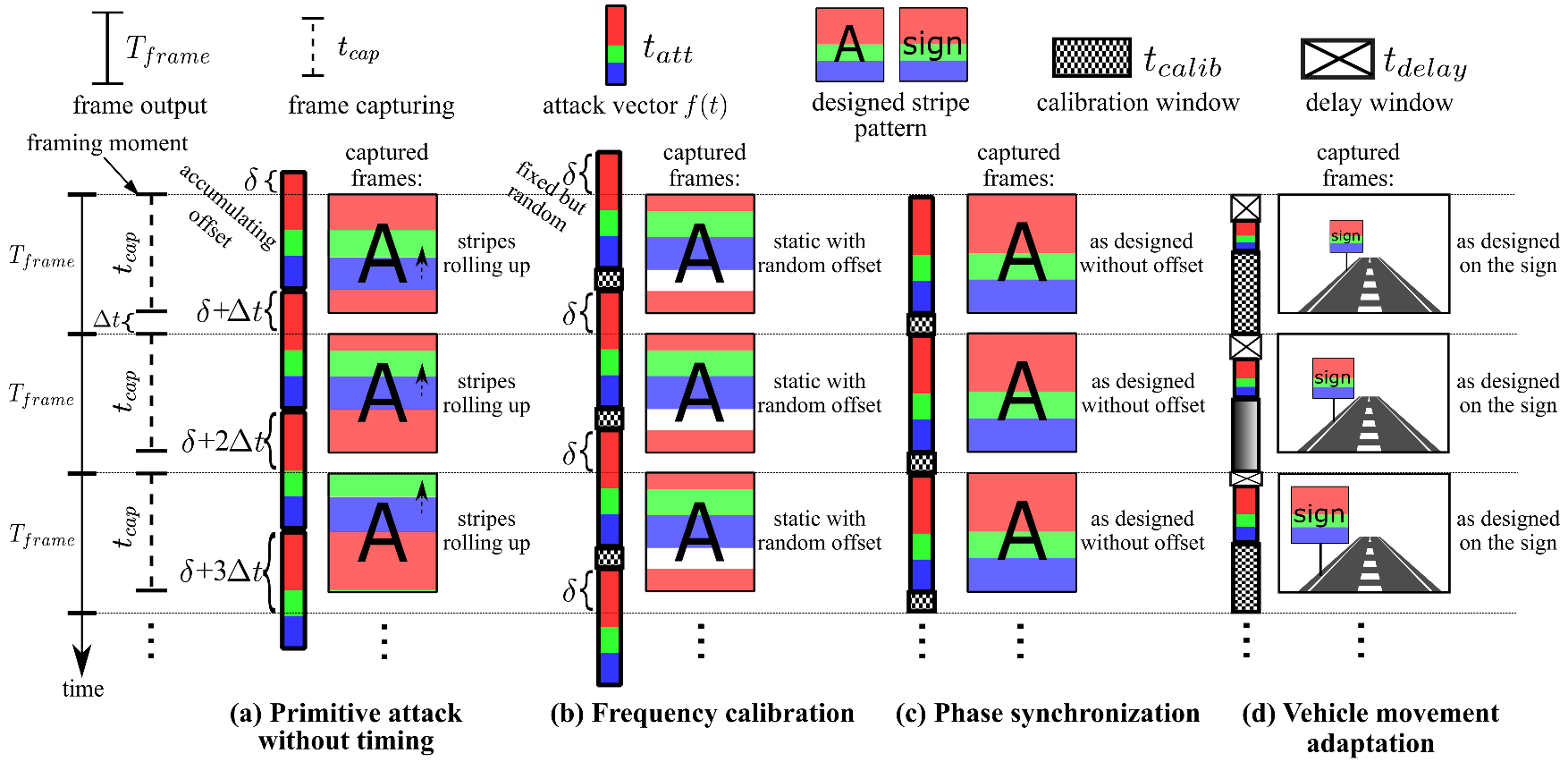}
  \vspace{-2em}
  \caption{Illustrations of the designs of attack timing control and vehicle movement adaptation.}
  \label{fig:stable_challenge}
\end{figure*}

This section analyzes two principles to achieve stable attack described in the introduction section, i.e., {\em attack timing control} and {\em vehicle movement adaptation}.

\subsection{Attack Timing Control}
\label{subsec:timing-control}

In this section, we analyze the simplified scenario described in \sect\ref{subsec:cvpr_paper}, i.e., the whole images in a frame sequence are classified. Figs.~\ref{fig:stable_challenge}a-c 
depict our analysis in this section. 
In reality, the vehicle classifies a sequence of image cropouts containing the traffic sign, as illustrated in Fig.~\ref{fig:stable_challenge}d. In \sect\ref{subsec:moving}, we will analyze how to deal with this real scenario.

To affect consecutive frames, the attacker needs to keep replaying the designed attack signal $f(t)$ where $t \in [0, t_{cap}]$ to control the LED.
Note that $T_{frame} \ge t_{cap}$ and we define $\Delta t \triangleq T_{frame} - t_{cap}$. In addition, we use $\delta$ to denote the time offset between the onset moment of the first play of $f(t)$ and the nearest camera's framing moment. A primitive attack, which continuously replays $f(t)$ back to back, accumulates $\Delta t$ over time on the offset between the replay's onset moment and the camera's framing moment. As illustrated in Fig.~\ref{fig:stable_challenge}a, the offset increases by $\Delta t$ for every frame. 
The resulting stripe pattern created by the attack rolls across the FoV over time (e.g., roll up in Fig.~\ref{fig:stable_challenge}a), leading to varying classification results.

To achieve a stable attack, the rolling needs to be avoided by {\em frequency calibration} such that the replay frequency is identical to the frame rate. This can be achieved by adding a calibration period $t_{calib} \triangleq T_{frame}-t_{cap}$ after each replay, as illustrated by the checkerboard squares in Fig.~\ref{fig:stable_challenge}b.
As such, the offset between the replay's onset moment and the camera's framing moment is fixed at $\delta$ over frames. 
The $\delta$ can take any value from $[-T_{frame}/2, T_{frame}/2]$, depending on the onset time of the attack. The resulted stripe pattern is stationary in the FoV, but the position offset is uncertain.
This uncertainty renders the attack untargeted.

If the attacker can further control its attack onset time such that $\delta = 0$ (which is called {\em phase synchronization}), the RSE-induced stripes will be identical to the designed pattern, as illustrated in Fig.~\ref{fig:stable_challenge}c. Hence, the victim's classification results over frames will be the target class $k$. 
To perform the phase synchronization, the attacker needs to obtain the framing moments, which can be sensed from the victim camera's magnetic emanation as we will detail in \sect\ref{subsec:att_sync}.

\subsection{Vehicle Movement Adaptation}
\label{subsec:moving}

The vehicle's traffic sign recognition pipeline only classifies the image cropout containing the detected traffic sign. Thus, only the RSE-induced stripes within the cropout affect the classification. As the position and size of the cropout in the FoV vary with time when the vehicle moves, the attack needs to adapt to the vehicle's movement. The adaptation logistics is analyzed as follows.

Assume that the upper edge of the cropout is at the $N_{up}$-th scanline counting from the top and the vertical dimension of the cropout is $N_{sign}$ scanlines. For ease of explanation, we analyze the case with phase synchronization. 
As illustrated in Fig.~\ref{fig:stable_challenge}d,
the attack can apply three time windows for timing control, i.e., {\em delay window}, {\em attack window}, and {\em calibration window}, represented by the crossed, colored and checkerboard squares, respectively.
The lengths of these three windows are: $t_{delay} = (N_{up}-1) \times t_{ro}$, $t_{att} = N_{sign} \times t_{ro} + t_{exp}$, and $t_{calib}=T_{frame}-t_{delay}-t_{att}$. The malicious LED flicking is performed within the attack window. When the victim vehicle moves, the $t_{delay}$, $t_{att}$, and $t_{calib}$ change over frames.
Therefore, the stripe pattern maintains as designed on the sign cropout  area that changes over frames.
For each frame, the LED control signal $f(t)$ over a time duration $t_{att}$ can be designed by solving $\argmin_{f(t)} \displaystyle \ell(\mathcal{M}(I_{cropout}), k)$, where $I_{cropout}$ is the image cropout affected by RSE. However, 
the high compute overhead of the online solving can easily breach the real-time requirement of the attack.
To simplify, we design an LED control signal $f_0(t)$ for a minimum attack window $t_{att0}$ during the offline stage. The $t_{att0}$ can be set according to the minimum size of the traffic sign in the FoV that can be detected. At run time, when $t_{att} \ge t_{att0}$, the $f(t)$ is obtained via scaling $f_0(t)$ up by $t_{att} / t_{att0}$ times,
and replayed during the attack window. When there is no phase synchronization, the replayed attack light signals can be filled into the calibration and delay windows to ensure that the perturbations appear on the traffic sign and avoid noticeable on-off flickering at the frame rate.

\section{GhostStripe Design}
\label{sec:system_design}

This section presents the design of GhostStripe. We first summarize the basic attack assumptions in \sect\ref{subsec:assumption}. Then, we overview the two versions of GhostStripe in \sect\ref{subsec:overview}. 
Then, the remaining three subsections present the approaches to attack signal optimization, vehicle movement adaptation, and phase synchronization, respectively.

\begin{figure*}[ht!]
  \centering
  \includegraphics[width=\linewidth]{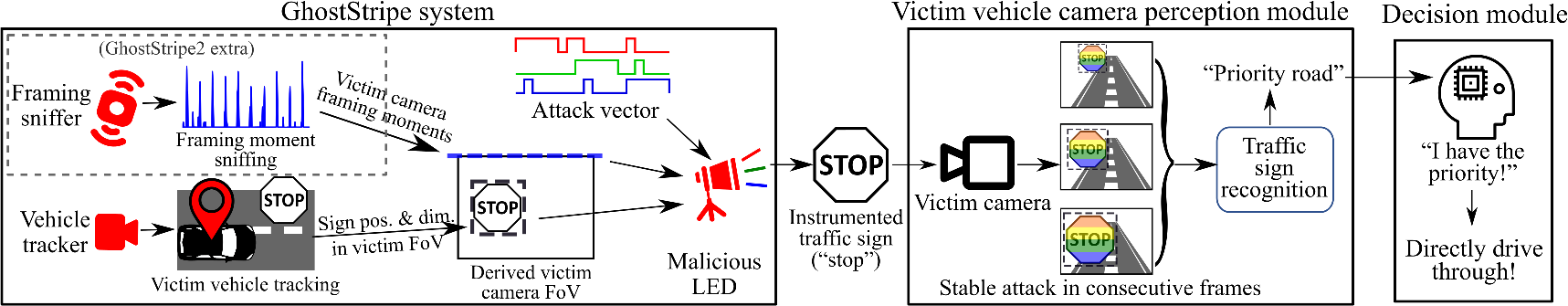}
  \caption{Overview of GhostStripe.}
  \label{fig:system_illustration}
\end{figure*}

\subsection{Basic Attack Assumptions}
\label{subsec:assumption}

The assumptions on the attacker are as follows: 
(1) The attacker can deploy a malicious LED to illuminate the traffic sign and a {\em vehicle tracker} to monitor the road section where the vehicles need to recognize the traffic sign. 
(2) The attacker needs to know the following fixed parameters of the victim vehicle's camera: focal length, sensor size, image resolution, and frame rate. 
These are commonly considered obtainable \cite{ji2021poltergeist,man2020ghostimage, wang2021can, kohler2021they, yan2022rolling}, e.g., from datasheets and reverse engineering on products.
For victim vehicles with auto-exposure feature enabled, 
the attacker can obtain the model on the relationship between $t_{exp}$ and ambient illumination and derive $t_{exp}$ at run time \cite{kohler2021they}. 
(3) The attacker has either white-box or black-box read access to the DNN used by the victim for traffic sign recognition. 
White-box means that the attacker knows the internals of the DNN (i.e., architectures and weights). Black-box means that the attacker only has the executable of the DNN and does not know its internals.
Obtaining DNNs might be harder but is assumed in all white-box \cite{eykholt2018robust,jing2021too, lovisotto2021slap,zhou2020deepbillboard,cao2021invisible, man2020ghostimage} and black-box \cite{ji2021poltergeist,lovisotto2021slap,man2020ghostimage,kohler2021they,yan2022rolling} attacks. It is possibly achievable from open codebases, by reverse engineering on products, or social engineering on manufacturers' employees.

\subsection{System Overview}
\label{subsec:overview}

We design two versions of GhostStripe, i.e., GhostStripe1 and GhostStripe2, with different 
requirements on the attack deployment to achieve untargeted and targeted stable attacks, respectively. 
GhostStripe1 maintains stationary adversarial stripes within the victim FoV by calibrating the LED flickering frequency and performs vehicle movement adaptation for real-time adjustment. 
It achieves untargeted attack.
On top of GhostStripe1, GhostStripe2 implements the phase synchronization to elimate the random offset  $\delta$. Therefore, the resulting adversarial stripe pattern remains same as designed and misleads the victim to produce the target class $k$. To achieve the phase synchronization, GhostStripe2 requires to clamp a sensor called {\em framing sniffer} onto the victim vehicle's camera power wire to sense the framing moments. Therefore, it targets a specific victim vehicle and controls the victim's traffic sign recognition results.

During the offline attack preparation phase, the attacker designs an LED control signal $f_0(t)$ for a minimum attack window $t_{att0}$ as described in \sect\ref{subsec:moving}. 
The workflow of GhostStripe during the online attack execution phase is illustrated in Fig.~\ref{fig:system_illustration}. The vehicle tracker tracks the real-time position of the victim vehicle and estimates the position and dimension of the traffic sign in the FoV of the victim vehicle's camera. In GhostStripe2, the framing sniffer senses the framing moments from the magnetic emanation of the camera power wire. 
Both the vehicle tracker and the framing sniffer continuously transmit their sensing results to the LED controller. Whenever the LED controller receives a report from either the vehicle tracker or the framing sniffer, it updates the attack signal and control parameters. Specifically, it scales up $f_0(t)$ to have $f(t)$ according to the dimension of the traffic sign and also determines the three time windows for attack timing control as illustrated in Fig.~\ref{fig:stable_challenge}d and \sect\ref{subsec:moving}.
The LED controller continuously replays the latest $f(t)$ with attack timing control.

\subsection{Attack Signal Optimization}
\label{subsec:attack_optim}

This section describes the generation of the minumum LED control signal $f_0(t)$.
To improve the robustness of the attack, $f_0(t)$ is obtained by solving
$\argmin_{f_0(t)} \mathbb{E}_{\phi} \left[ \ell(\mathcal{M}(I_{sign}^{\phi}), k)\right]$, 
where $\phi$ represents the uncontrollable offset in terms of the number of scanlines;
$I_{sign}^{\phi}(u,v)=I_{sign,amb}(u,v) + I_{sign,att}(u,v) \cdot g(v+\phi)$ is the image cropout containing the traffic sign;
$I_{sign,amb}(u,v)$ and $I_{sign,att}(u,v)$ are the corresponding image cropouts from $I_{amb}(u,v)$ and $I_{att}(u,v)$ defined in \sect\ref{subsec:cvpr_paper}. 
For GhostStripe1, since there is no control on the offset, we sample $\phi$ uniformly from $[0, N_{sign}]$  to evaluate the mathematical expectation of the objective function; 
for GhostStripe2, as the phase synchronization can largely reduce the offset, we sample $\phi$ uniformly from a narrow range of $[-0.1N_{sign}, 0.1N_{sign}]$, where the multiplier 0.1 is empirically chosen.

\textbf{White-box optimization.}
Since the analytical model of the rolling shutter as described in \sect \ref{subsec:cvpr_paper} is differentiable, $f_0(t)$ can be obtained by gradient-based methods.
We use Projected Gradient Descent (PGD) \cite{madry2018towards}, which iteratively perturbs input data towards maximizing the loss function while maintaining the perturbations within a bounded range, i.e., $f_{0}(t) \in [0, 1]$. By iteratively adjusting the $f_{0}(t)$ based on the attainable internal gradients, PGD can efficiently optimize the $f_{0}(t)$ against the victim model.

\textbf{Black-box optimization.}
We implement Bayesian Optimization (BO) \cite{mockus2005bayesian,bayesianopttool}, which is a strategy for global optimization of black-box functions. 
It involves a Bayesian statistical model and an acquisition function. The statistical model generates a Bayesian posterior probability distribution to approximate the objective function, updated with each new query. Subsequently, this posterior distribution is utilized to construct the acquisition function, determining the next query point. 
With black-box access, we query the model with attacked images $I(u,v)$, and obtain prediction classes and confidence outputs. This allows BO to iteratively refine $f_{0}(t)$ based on the model's responses.
Since BO is suitable for problems in low cardinality (typically, lower than $30$), we reduce the cardinality of $f_0(t)$ by restructuring each color channel $f^c_0(t)$ as a vector of length $q$. Each element lasts for a time period $t_{att0}/q$. This limits BO's search space dimension to $3 \times q$ for the three color channels of $f(t)$. In terms of perturbation appearance, the final perturbation consists of $q$ stripes with equal vertical length, in contrast to the stripes in the white-box setting that are on a scanline-wise basis. In our implementation, we experimentally choose $q$ from 5 to 10 and use the one that yields the best attack effectiveness.

\subsection{Locating Traffic Sign in Camera FoV}
\label{subsec:fov_estimation}
\begin{figure}
  \centering
  \begin{subfigure}[t]{0.43\linewidth}
    \includegraphics[width=\linewidth]{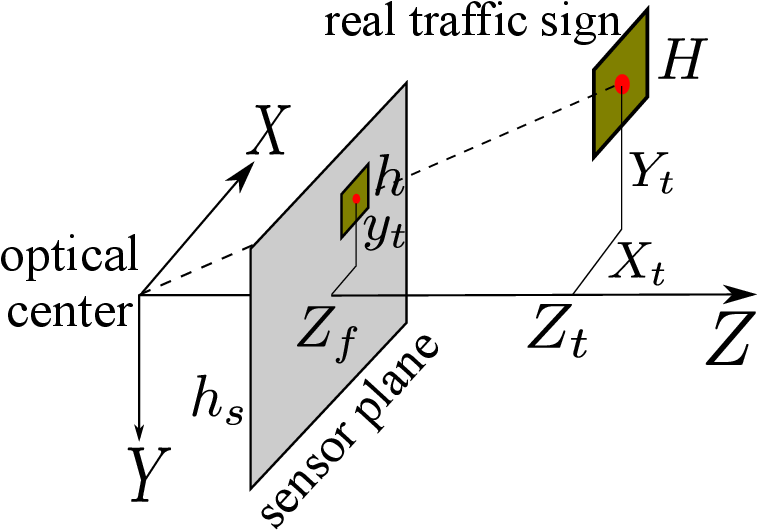}
    \caption{Prospective projection model.}
    \label{sfig:perspective_projection_3d}
  \end{subfigure}
  \hfill
  \begin{subfigure}[t]{0.51\linewidth}
  \centering
    \includegraphics[width=\linewidth]{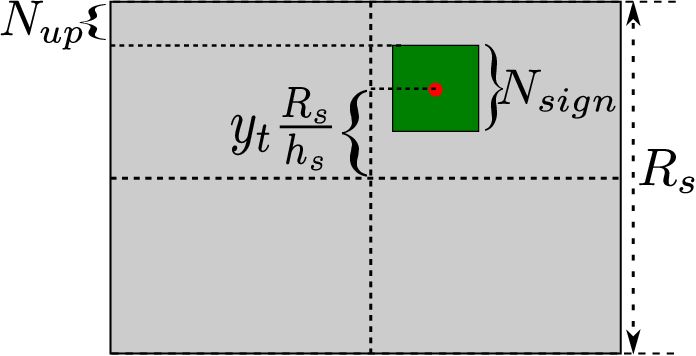}
    \caption{Traffic sign's projection in the sensor plane.}
    \label{sfig:image_plane}
  \end{subfigure}
  \caption{Estimation of the traffic sign's vertical position and size in the captured image.}
  \label{fig:victim_view_derivation}
\end{figure}

This section presents the approach to estimating the traffic sign's vertical position and size in the victim vehicle camera's FoV. Its principle based on the prospective projection model is described as follows.
Fig.~\ref{sfig:perspective_projection_3d} shows an {\em ego coordinate system} originating from the victim camera's optical center, where
the $X$- and $Y$-axes define the camera sensor plane, and the $Z$-axis is the optical axis perpendicular to the camera sensor plane. Let $(X_t, Y_t, Z_t)$ and $H$ denote the coordinates of the traffic sign's center and the vertical dimension of the traffic sign, respectively.
Let $Z_{f}$ and $h_s$ denote the victim camera's focal length and the vertical dimension of the camera sensor.
From Fig.~\ref{sfig:perspective_projection_3d},
the vertical position and size of the traffic sign's projection on the sensor plane are $y_t = Z_f \frac{Y_t}{Z_t}$ and $h = Z_f \frac{H}{Z_t}$, respectively.
Denoting by $R_s$ the total number of the camera's scanlines.
A unit length of the sensor plane's vertical dimension corresponds to $\frac{R_s}{h_s}$ scanlines. Fig.~\ref{sfig:image_plane} shows the sensor plane and the projection of the traffic sign. The projection's vertical size and position in scanlines can be derived as $N_{sign} = h \frac{R_s}{h_s} = Z_f \frac{H}{Z_t} \frac{R_s}{h_s}$
and $N_{up}= \frac{1}{2} R_s - y_t \frac{R_s}{h_s} - \frac{1}{2} N_{sign} = \frac{1}{2} R_s - (Y_t+\frac{1}{2}H)\frac{Z_f R_s}{Z_t h_s}$. Note that the values of $Z_f$, $h_s$, and $R_s$ are available from the camera's datasheet; the traffic sign size $H$ can be measured by the attacker.

From the above analysis, to estimate $N_{sign}$ and $N_{up}$, the attacker needs to obtain $Y_t$ and $Z_t$.
If the victim vehicle is on a flat road section, $Y_t$ is the altitude difference of the traffic sign and the vehicle camera. The traffic sign's altitude can be measured by the attacker; the vehicle camera's altitude can be obtained from the vehicle specification or measured by the attacker as well. The $Z_t$ is the horizontal distance between the victim vehicle and the traffic sign, which can be obtained by localizing the victim vehicle in real time. 
With $Z_t$, the updated $N_{sign}$ and $N_{up}$ are used for vehicle movement adaptation.

The victim camera's pitch angle and road gradient can affect the traffic sign's vertical position in the camera's FoV. The pitch angle can be obtained from the vehicle specification or measured. The road gradient can be measured in advance to optimize the attack.
Both can be factored in when determining $Y_t$.

\subsection{Phase Synchronization}
\label{subsec:att_sync}
\begin{figure}
  \centering
  \includegraphics[width=0.9\linewidth]{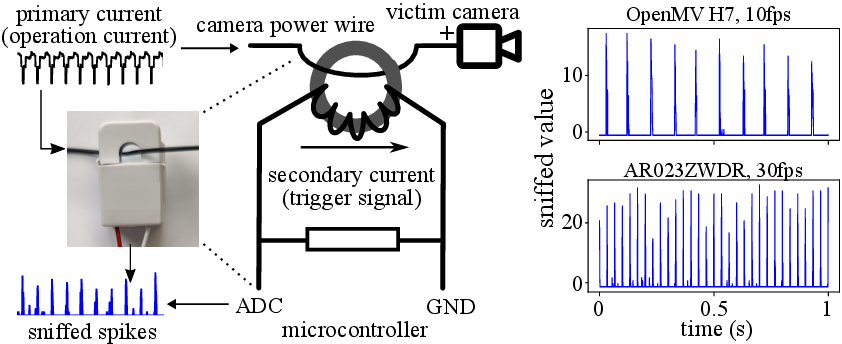}
  \vspace{-1em}
  \caption{Framing sniffer and measurement traces.}  
  \label{fig:sensor_circuit}
\end{figure}

\begin{figure}
  \centering
  \begin{subfigure}[t]{.476\linewidth}
    \includegraphics[width=\linewidth]{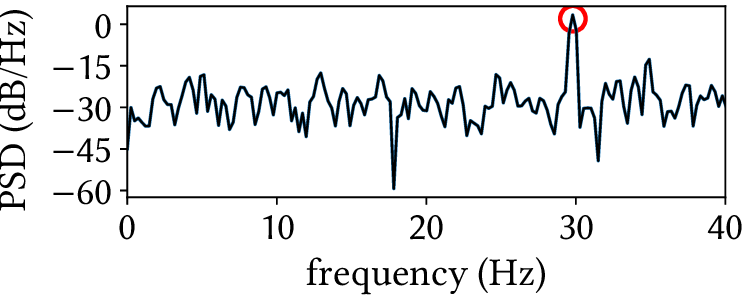}
    \caption{Logitech V-U0018, $30\,\text{fps}$}
    \label{sfig:logitech_sniff}
  \end{subfigure}
  \begin{subfigure}[t]{.476\linewidth}
    \includegraphics[width=\linewidth]{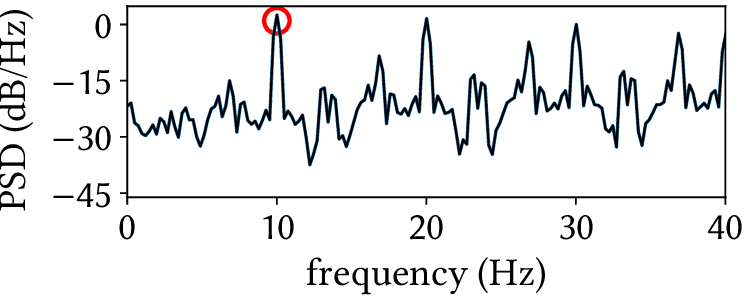}
    \caption{OpenMV H7, $10\,\text{fps}$}
    \label{sfig:openmv_sniff}
  \end{subfigure}
  \hfill
  \begin{subfigure}[t]{.476\linewidth}
    \includegraphics[width=\linewidth]{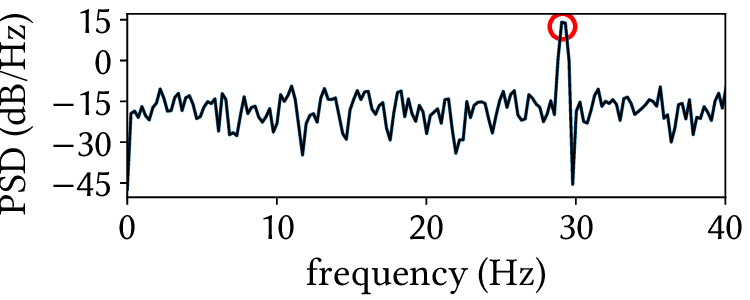}
    \caption{AR1820HS, $29\,\text{fps}$}
    \label{sfig:arducam_sniff}
  \end{subfigure}
  \begin{subfigure}[t]{.476\linewidth}
    \includegraphics[width=\linewidth]{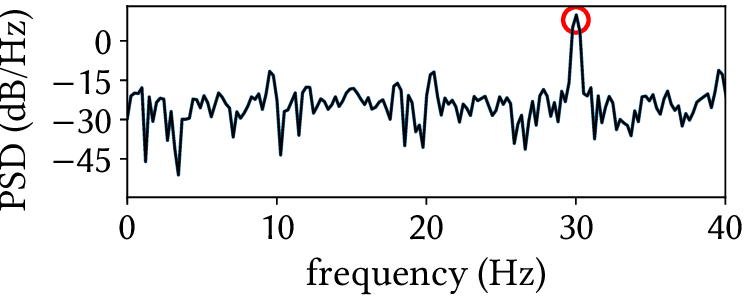}
    \caption{AR023ZWDR, $30\,\text{fps}$}
    \label{sfig:licam_sniff}
  \end{subfigure}
  \vspace{-1em}
  \caption{PSDs of the magnetic emissions of cameras.}
  \label{fig:sniff_time_freq_profile}
\end{figure}

This section presents how GhostStripe2 senses the victim camera's framing moments to achieve phase synchronization.
The internal operations of a camera may create variations in the camera's current draw and the resulting magnetic emanation. We investigate whether the emanation provides salient characteristics for inferring framing moments of four off-the-shelf cameras:
Logitech V-U0018,
OpenMV H7,
Arducam AR1820HS,
and Leopard Imaging AR023ZWDR.
The last one is the camera product in Baidu Apollo's hardware reference design \cite{apollohardware}.
The frame rates of these cameras are $30$, $10$, $29$, and $30\,\text{fps}$, respectively.
To sense the magnetic emanation, as shown in Fig.~\ref{fig:sensor_circuit}, we integrate a YHDC SCT-006 split-core current transducer with a $330\,\Omega$ resistor and sample the voltage over the resistor using an Arduino Due. The current transducer is clamped onto the camera's power wire. The current in the wire generates a magnetic field concentrated at the magnetic split-core, which further induces a secondary current in the winding and then a voltage over the resistor.
Fig.~\ref{fig:sensor_circuit} also shows the measurement traces for two cameras. We can see periodic time-domain spikes. The interval between two spikes is about $T_{frame}$. Fig.~\ref{fig:sniff_time_freq_profile} shows the power spectral densities (PSDs) of the measurement traces for the four cameras.
The highest PSD peak appears at the camera's frame rate. These results suggest that the time-domain spikes may be indicative of framing moments.

\begin{figure*}[h!]
  \centering
  \begin{subfigure}{0.57\linewidth}
    \centering
    \includegraphics[width=\linewidth]{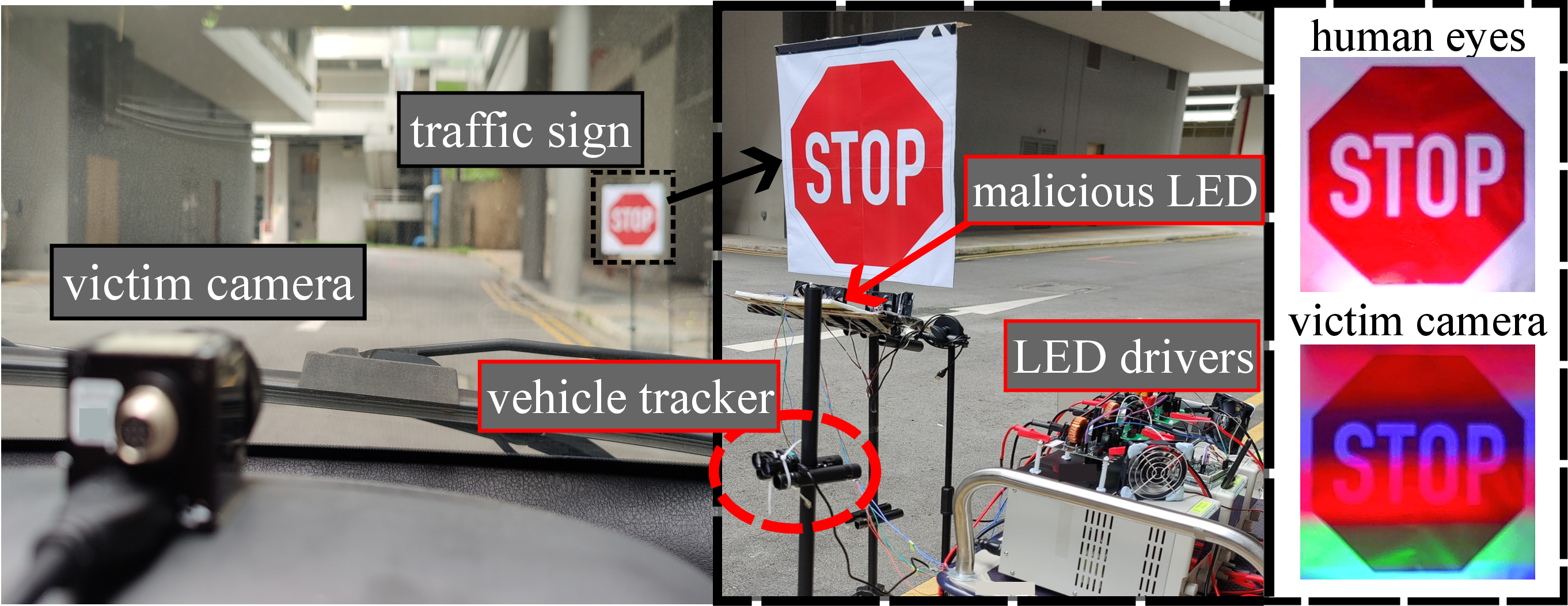}
    \vspace{-1.5em}
    \caption{Outdoor testbed.}
    \label{sfig:outdoor_testbed}
  \end{subfigure}
  \begin{subfigure}{0.24\linewidth}
    \includegraphics[width=\linewidth]{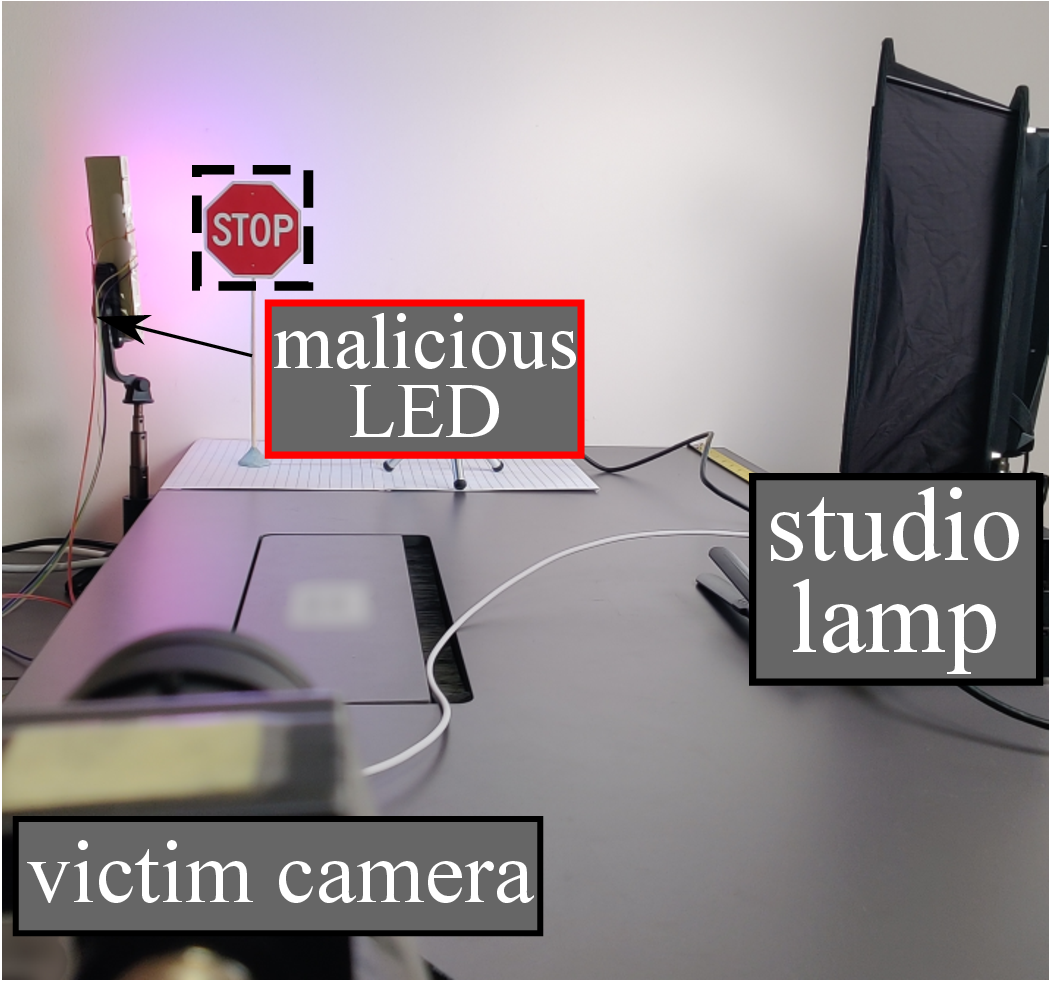}
    \vspace{-1.5em} 
    \caption{Lab testbed.}
    \label{sfig:testbed_outlook}
  \end{subfigure}
  \begin{subfigure}{0.165\linewidth}
    \includegraphics[width=\linewidth]{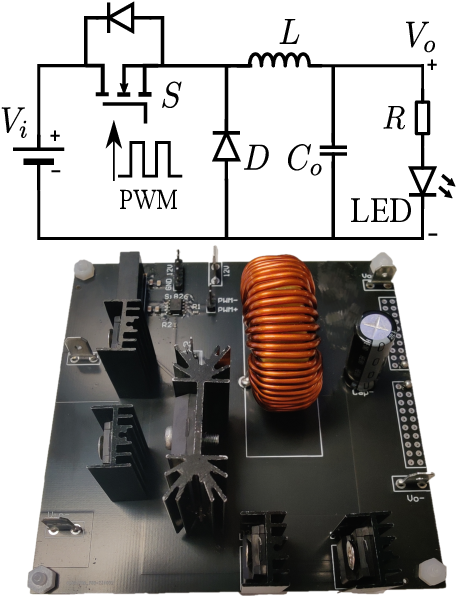}
    \vspace{-1.5em}
    \caption{LED driver.}
    \label{sfig:led_attacker}
  \end{subfigure}
  \vspace{-1em}
  \caption{Testbed setups and the LED driver.}
  \label{fig:testbed_setup}
\end{figure*}

\begin{figure}
  \centering
  \begin{subfigure}[t]{0.57\linewidth}
    \includegraphics[width=\linewidth]{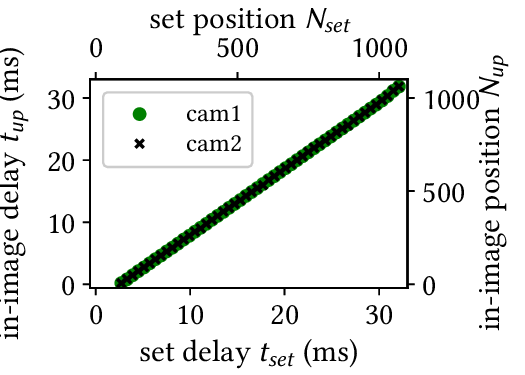}
    \caption{Set delay vs. in-image delay on two AR023ZWDR cameras.}
    \label{sfig:sync_set_vs_actual}
  \end{subfigure}
  \hfill
  \begin{subfigure}[t]{0.365\linewidth}
    \includegraphics[width=\linewidth]{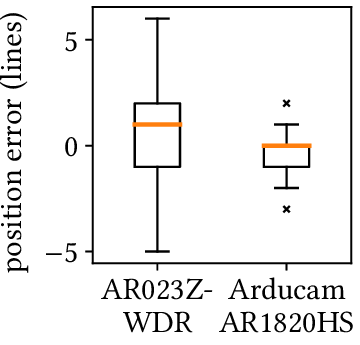}
    \caption{Position error on two camera models.}
    \label{sfig:sync_mapping_error}
  \end{subfigure}
  \vspace{-1em}
  \caption{Phase synchronization performance. }
  \label{fig:sync_analysis}
\end{figure}

The sniffer uses a threshold to detect the time-domain spikes. To wirelessly trigger the LED controller with the detected spikes, we use two Nordic nRF24L01+ transceivers operating in the $2.4\,\text{GHz}$ ISM band. Upon detecting a spike, the sniffer transmits a packet to the LED controller, which then prompts the replay of the light signals upon packet detection.

We design experiments to investigate how to use the time-domain spikes to perform phase synchronization.
We present the experiment results for two AR023ZWDR cameras, where $t_{ro}=30\,\mu\textit{s}$ and $t_{exp}$ is set to $1\,\text{ms}$.
In a dark room, we light up the LED after a {\em set delay} $t_{set}=(N_{set}-1) \times t_{ro}$ from each detected spike.
The LED is on for a short period to
form a bright stripe in the dark background of the camera's FoV. We find the top lighten scanline and extract its vertical coordinate $N_{up}$ from the FoV top. We also compute the actual \emph{in-image delay} as $t_{up} = (N_{up}-1) \times t_{ro}$. If the spike precisely indicates the framing moment, we should have $N_{up} = N_{set}$, i.e., the vertical position of the stripe can be precisely controlled at $N_{set}$.
Fig.~\ref{sfig:sync_set_vs_actual} shows the $t_{up}$ versus $t_{set}$ and $N_{up}$ versus $N_{set}$ when we vary $N_{set}$. The results obtained on two separate AR023ZWDR cameras are shown.
Analysis on the results shown in Fig.~\ref{sfig:sync_set_vs_actual} suggests that $N_{up} - N_{set}$ is non-zero but the $t_{set}$-versus-$N_{up}$ relationship shows high consistency across the two cameras. Therefore, by using this relationship, we can choose the $t_{set}$ value according to the desired $N_{up}$ to control the LED. We evaluate the error between the desired $N_{up}$ and the actual $N_{up}$ on a camera when the $t_{set}$ is determined by the $t_{set}$-versus-$N_{up}$ relationship obtained on the other camera.
Fig.~\ref{sfig:sync_mapping_error} shows the results.
The maximum error is 6 scanlines, which is merely 0.55\% of the vertical resolution of the camera (i.e., 1,088 scanlines). We also profile the $t_{set}$-versus-$N_{up}$ on an Arducam AR1820HS camera and evaluate the $N_{up}$ control error on a different Arducam AR1820HS. The maximum error is 3 scanlines. The above results show that precise phase synchronization can be achieved by using the sensing results of the framing sniffer.

\section{GhostStripe Implementation}
\label{sec:testbed_and_prototype}

\subsection{Testbed Setups}
\label{subsec:testbed}

\textbf{Victim camera:} 
We use Leopard Imaging AR023ZWDR as the victim camera. It is the default main camera in Baidu Apollo's hardware reference design \cite{apollohardware}. 
It is built upon the ONSEMI AR023Z rolling shutter-based image sensor with a size of $5.78\,\text{mm} \times 3.26\,\text{mm}$, $1928 \times 1088$ active pixels. Each scanline has a readout time $t_{ro}$ of $30\,\mu\text{s}$. Its focal length is $12\,\text{mm}$ .

\textbf{Outdoor testbed:} 
We use a real road section and a real car, as shown in Fig.~\ref{sfig:outdoor_testbed}.
We deploy most common traffic signs \cite{mostcommontrafficsigns} including  ``stop'',``yield'', and ``speed limit'' with size and altitude conforming to the Manual on Uniform Traffic Control Devices (MUCTD) \cite{MUTCD}.
We mount the victim camera under the front windshield of the car. The sign-car distance for the camera to perceive the whole sign is from $10\,\text{m}$ to $32\,\text{m}$.

\textbf{Lab testbed:}  We build a lab testbed in 1:10 scale as shown in Fig.~\ref{sfig:testbed_outlook} to simulate a road section. The total length of the testbed is $3.6\,\text{m}$.
We deploy common signs including ``stop'', ``yield'', and ``speed limit''.
To control ambient illumination condition, we set up two studio lamps with tunable intensity to project light onto the testbed. The color temperature of the lamps is $5600\,\text{K}$, which is similar to normal sunlight.
This lab setup allows us to isolate the impact of uncontrollable environment factors and provide better understanding of the impacts of several factors on GhostStripe.

\textbf{Traffic sign recognition models.} We integrate the YOLO object detector \cite{redmon2016you} and an AlexNet-based 8-layer convolutional neural network traffic sign classifier.
We train the classifier on the German Traffic Sign Recognition Benchmark (GTSRB) dataset \cite{stallkamp2011german}, which contains over 50,000 image samples in 43 classes. The trained model achieves a 95.35\% accuracy on the GTSRB testing set.
When we test the trained model with numerous video frames taken for the signs deployed in our testbeds in the absence of attack, it achieves 100\% accuracy under various camera poses, distances, and illumination conditions considered in our experiments.

\subsection{GhostStripe Implementation}
\label{subsec:prototype}

With the capabilities presented in \sect\ref{subsec:fov_estimation} and \sect\ref{subsec:att_sync}, we implement GhostStripe by following the workflow presented in \sect\ref{subsec:overview}. 
The replay of a given $f(t)$ is implemented by pulse-width modulation (PWM) for the LED's power supply using an Arduino Due.
We integrate 30 and 4 Marktech XM-L RGB LED units to emit the attack light in the outdoor and lab testbeds, respectively. 
To achieve higher attack light intensity for outdoor implementation, we customize three buck converters for the three color channels respectively to form an LED driver.
Each converter takes the PWM signal of a color channel from the Arduino Due to regulate the high input voltage drawn from a direct current power supply, and drives the LEDs to emit attack light. Fig.~\ref{sfig:led_attacker} shows the design schematic and the fabricated LED driver.

\begin{figure}
  \centering
  \begin{subfigure}[t]{0.60\linewidth}
    \includegraphics[width=\linewidth]{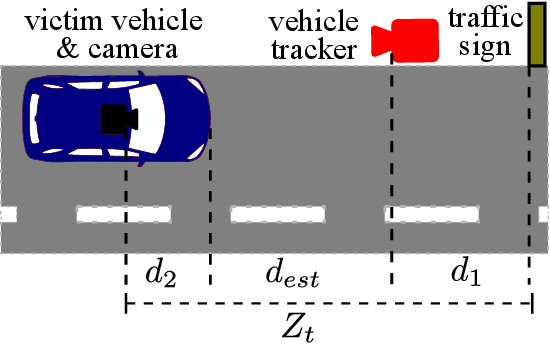}
    \caption{Layout.}
  \label{fig:watcher_camera_placement}
  \end{subfigure}
  \hfill
  \begin{subfigure}[t]{0.36\linewidth}
    \includegraphics[width=\linewidth]{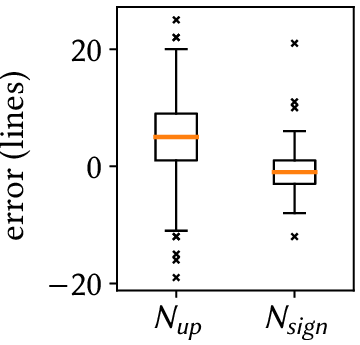}
    \caption{Estimation errors.}
    \label{sfig:fov_testbed_error}
  \end{subfigure}
  \vspace{-1em}
  \caption{Vehicle localization and FoV estimation.}
  \label{fig:testbed_fov_estimation}
\end{figure}

For the vehicle tracker, we implement an essential victim vehicle localization function. As shown in Fig.~\ref{fig:watcher_camera_placement}, the vehicle tracker, which is based on a LightWare SF30/C LiDAR rangefinder, is placed on the road side facing the upcoming traffic, measuring the distance $d_{est}$ to the vehicle in real time.
We measure the distance between the traffic sign and the vehicle tracker (denoted by $d_1$), the distance between the victim camera and the vehicle front surface (denoted by $d_2$), the altitudes of the traffic sign and the victim camera (denoted by $Y_{sign}$ and $Y_{cam}$). Thus, in the victim camera's ego coordinate system, the $Y_t$ and $Z_t$ needed by GhostStripe are given by $Y_{t}=Y_{sign} - Y_{cam}$ and $Z_{t} = d_{est} + d_{1} + d_{2}$. 
As shown in Fig.~\ref{sfig:fov_testbed_error}, the resulted $N_{up}$ and $N_{sign}$ estimates have errors less than 20 scanlines (i.e., 1.8\% of the camera's vertical resolution).

\section{Evaluation}
\label{sec:evaluation}

We evaluate GhostStripe's attack effectiveness by testing it against the camera on a moving vehicle in the outdoor testbed. Additionally, we examine the effects of several important factors using the lab testbed. Throughout this section, we use the abbreviation {\bf \emph{GS}} to refer to GhostStripe.

\subsection{Evaluation Methodology}

\subsubsection{Evaluation metrics}
We use the following metrics to characterize attack effectiveness:
(1) \textbf{Misclassification rate (MR)}: 
MR is the ratio of frames where the traffic sign is incorrectly identified as a non-ground-truth class, divided by the total number of frames.
(2) \textbf{Primary misclassification class rate (PMCR)}: 
The primary misclassification class is defined as the most frequently misclassified class when \emph{GS1} is deployed, or the targeted class when \emph{GS2} is deployed.
PMCR is the ratio of frames where the traffic sign is misclassified as the primary misclassification class to the total number of frames.
(3) \textbf{Entropy}:
We employ Shannon entropy to quantify the randomness of classification results within a time window. In this section, we compute the entropy values within $1.5\,\text{s}$ time windows, adopted from the window size for decision making used in Baidu Apollo's traffic light recognition. Lower entropy values signify increased stability in classification results.

\subsubsection{Baselines}
We employ the following baseline attack approaches.
(1) The {\bf \emph{Random}} approach employs randomly appeared colored stripes;
(2) The {\bf \emph{Primitive}} approach \cite{sayles2021invisible} generates the colored stripes with an offset-robust design which is also used in \emph{GS1} as described in \sect\ref{subsec:attack_optim}, without timing control for stable attack.
(3) {\bf \emph{GS2-still}} approach is a variant of \emph{GS2} that is designed for a specific victim location and does not employ vehicle movement adaptation. This baseline is used to understand the contribution of vehicle movement adaptation to the attack performance.

\begin{figure*}[ht!]
  \begin{subfigure}[t]{0.32\linewidth}
    \includegraphics[width=\linewidth]{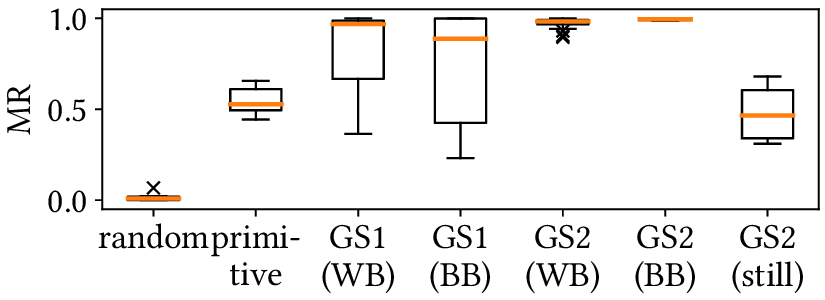}
    \vspace{-1.5em}
    \caption{MR.}
    \label{sfig:dynamic_mr}
  \end{subfigure}
  \begin{subfigure}[t]{0.32\linewidth}
  \centering
    \includegraphics[width=\linewidth]{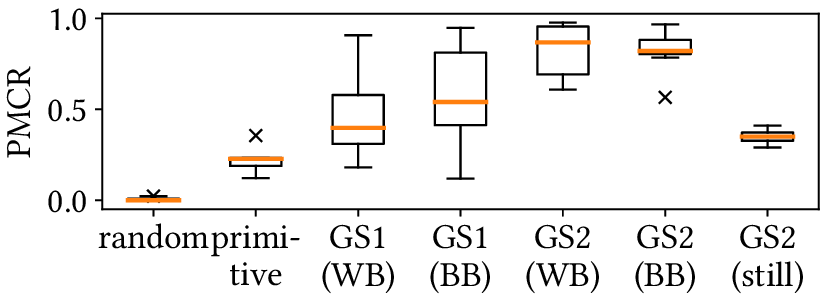}
    \vspace{-1.5em}
    \caption{PMCR.}
    \label{sfig:dynamic_pmcr}
  \end{subfigure}
  \begin{subfigure}[t]{0.32\linewidth}
  \centering
    \includegraphics[width=\linewidth]{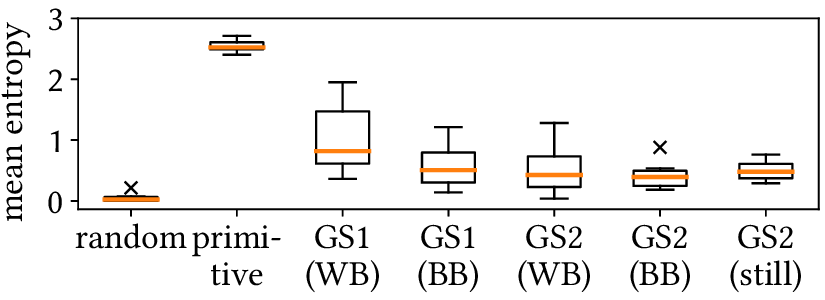}
    \vspace{-1.5em}
    \caption{Mean entropy.}
    \label{sfig:dynamic_entropy}
  \end{subfigure}
  \caption{Comparison with baseline methods. Abbreviations: ``WB'' for ``white-box'', and ``BB'' for ``black-box''.}
  \label{fig:dynamic_baselines}
\end{figure*}
\subsection{Evaluation on Outdoor Testbed}
\label{subsec:eval_outdoor}

\subsubsection{Impact on detection}
\label{subsubsec:impact_on_detection}

We assess \emph{GS}'s impact on traffic sign detection (i.e., the step prior to recognition). We measure the Intersection over Union (IoU) of the  detection results obtained at different vehicle-sign distances.
The detector achieves consistently high IoU of about 0.94 during the \emph{GS} attack. When using these detection results to select cropouts from clean images when the attack is temporarily switched off, all cropouts are correctly classified.
Thus, \emph{GS} has negligible impact on the traffic sign detector.

\subsubsection{Overall attack performance}

We study the effectiveness of GS against a moving vehicle using the most representative sign ``stop'' as an example. In this subsection, we plan the attack based on a camera exposure time of $1/1000\,\text{s}$. First, we present the results obtained during the offline attack optimization phase. \emph{Random} can rarely deviate the classification results from the ground truth. 
With \emph{Primitive} and \emph{GS1} which share the same attack signals optimized for the whole offset range, the untargeted attack across all the offsets succeeds at a rate of 87.2\% in the white-box setting, and 81.1\% in the black-box setting.
For \emph{GS2}, we choose the ``priority road'' sign as the target class, which is semantically conflicting with the stop sign.
\emph{GS2} achieves 100\% targeted attack success rate, in both white-box and black-box settings. 

Then, we test the attacks on the testbed during normal daytime hours (9 am to 5pm) under partly cloudy weather conditions. In this set of experiments, we drive the vehicle along the road section at a speed of around $10\,\text{km/h}$ and record video footage containing the traffic sign under attack.
Fig.~\ref{fig:dynamic_baselines} provides a summary of the overall attack performances for different methods.
\emph{Random} is ineffective, as the MR and PMCR are both almost zero. 
\emph{Primitive} achieves a mean MR of 54.5\% and PMCR of 22.4\%. However, the mean entropy is high at 2.55. These results suggest that \emph{Primitive} induces unstable classification results within each $1.5\,\text{s}$ window due to the varied stripe patterns on the sign cropout across frames.

Both \emph{GS1} and \emph{GS2} perform effectively, regardless of whether they are generated with white-box or black-box (indicated as ``WB'' and ``BB'' in Fig.~\ref{fig:dynamic_baselines}, respectively) DNN knowledge. \emph{GS2} exhibits the highest performance in targeted attacks, achieving mean PMCRs of 83.2\% under the white-box setting, and 82.4\% under the black-box setting. Here the PMCRs of white-box setting show more variation than black-box setting. This is likely due to the varying testing conditions across trials. While the white-box attack requires more information, its main benefit lies in optimization efficiency. After successful training, white-box attack is not necessarily more effective than black-box at runtime, as effectiveness depends on testing conditions.
\emph{GS1} demonstrates a high success rate in untargeted attacks, with mean and median MRs of 81.5\% and 96.8\% under the white-box setting and 73.4\% and 88.7\% under the black-box setting.
Note that the primary misclassification class in \emph{GS1} may vary across trials as different perturbation offsets may result in different classes.
Although the PMCRs of \emph{GS1} hover at around 50\%, which are lower than \emph{GS2}, they are still higher than other methods. 
The relatively low PMCR of {\em GS1} compared with {\em GS2} is explained as follows. During the {\em GS1}'s offline attack signal optimization, the vertical offset $\phi$ is sampled from a wide range. As such, adjacent offsets may not result in the same class.
Consequently, at runtime, when slight misalignments occur between the designed stripes and the sign cropout in the victim FoV, the misclassification results may vary. However, the relatively stable stripe pattern in \emph{GS1} still contributes to overall attack stability, as indicated by the slight entropy increase compared with \emph{GS2}.

\begin{figure}
  \centering
  \begin{subfigure}[t]{0.325\linewidth}
    \includegraphics[width=\linewidth]{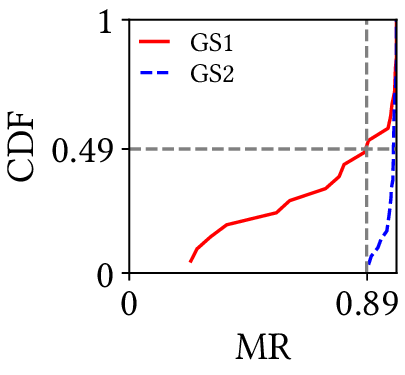}
    \caption{MR.}
    \label{sfig:gs1_vs_gs2_mr}
  \end{subfigure}
  \hfill
  \begin{subfigure}[t]{0.325\linewidth}
  \centering
    \includegraphics[width=\linewidth]{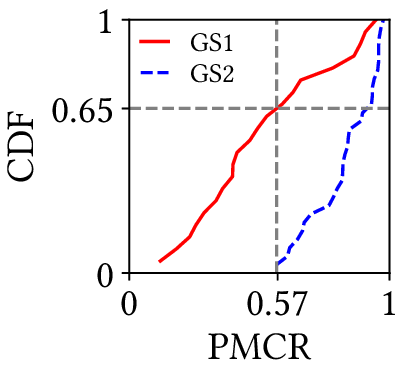}
    \caption{PMCR.}
    \label{sfig:gs1_vs_gs2_pmcr}
  \end{subfigure}
  \hfill
  \begin{subfigure}[t]{0.325\linewidth}
  \centering
    \includegraphics[width=\linewidth]{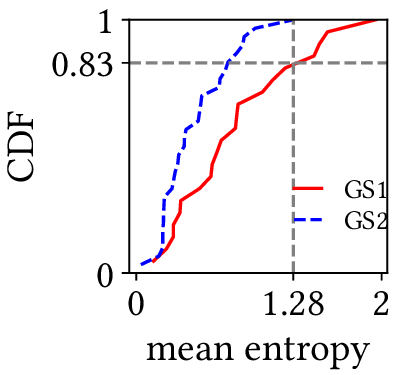}
    \caption{Entropy.}
    \label{sfig:gs1_vs_gs2_entro}
  \end{subfigure}
  \vspace{-1em}
  \caption{Comparison between GS1 and GS2.}
  \label{fig:cdfs_gs1_gs2}
\end{figure}

We also compare \emph{GS1} and \emph{GS2} in Fig.~\ref{fig:cdfs_gs1_gs2}. 
For \emph{GS2}, the minimum MR and PMCR, and maximum mean entropy are 89.5\%, 56.6\%, and 1.28. On \emph{GS1}'s cumulative distribution function (CDF) curves, the corresponding probabilities are 49\%, 65\%, and 83\%, as illustrated in Fig.~\ref{sfig:gs1_vs_gs2_pmcr} and \ref{sfig:gs1_vs_gs2_entro}. 
The interpretation of these results are as follows. In terms of MR, \emph{GS1} can perform no worse than \emph{GS2} in $100\%-49\%=51\%$ cases for spoofing traffic sign to any other class during one run. In terms of PMCR, \emph{GS1} can perform no worse than \emph{GS2} in $100\%-65\%=35\%$ cases for spoofing traffic sign to a primary misclassification class during one run, although this class is not controllable. In terms of entropy within each time window, \emph{GS1} can perform no worse than \emph{GS2} in 83\% cases.

\emph{GS2-still} achieves 48.2\% mean MR, 35.3\% PMCR, and 0.50 mean entropy. The performance drop compared with \emph{GS2} is because when the stripes fall on the traffic sign in the FoV, the attack is targeted; otherwise, the results are unpredictable. This shows the benefit of continuous vehicle tracking and movement adaptation, for enhancing attack effectiveness compared with a static attack targeting a specific position.

\begin{figure}
  \centering
  \includegraphics[width=\linewidth]{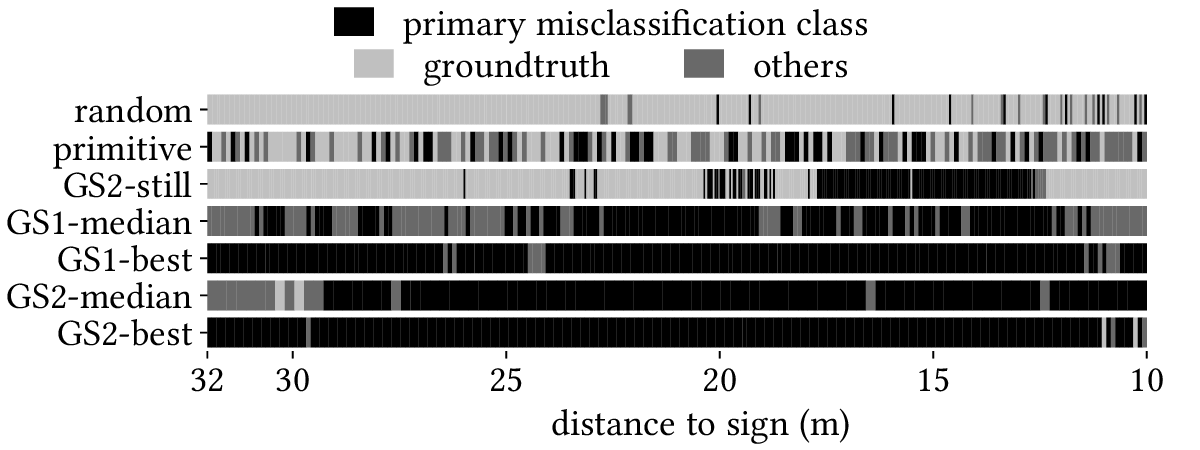}
  \vspace{-2em}
  \caption{Example of attack results on the consecutive frames when the vehicle passes the road section.}
  \label{fig:dynamic_frames}
\end{figure}

\subsubsection{Visualization of attack effectiveness.}
We illustrate the attack effectiveness of the attack results by drawing the classification results when the vehicle drives through the road section, as shown in Fig.~\ref{fig:dynamic_frames}. 
\emph{GS1-median} and \emph{GS2-median} denote the result traces in the runs where \emph{GS1}'s and \emph{GS2}'s PMCRs are around their respective median levels.
\emph{GS1-best} and \emph{GS2-best} denote the best result traces of \emph{GS1} and \emph{GS2} in all runs.
Both \emph{GS1} and \emph{GS2} achieve relatively stable attack effectiveness. 
In the best cases, \emph{GS1} and \emph{GS2} can achieve attack success rates of over 94\% and 97\%, respectively, in misleading the victim to the primary misclassification class stably.
In contrast, baseline attack methods show ineffectiveness and/or result randomness.

\begin{figure}
  \centering
  \begin{minipage}[t]{0.48\linewidth}
    \centering
    \includegraphics[width=\textwidth]{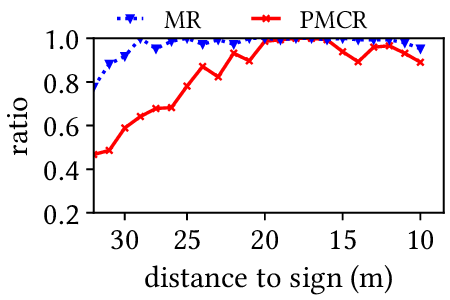}
    \vspace{-2em}
    \caption{Impact of sign-vehicle distance.}
    \label{fig:impact_of_distance}
  \end{minipage}
  \hfill
  \begin{minipage}[t]{0.48\linewidth}
    \centering
    \includegraphics[width=\textwidth]{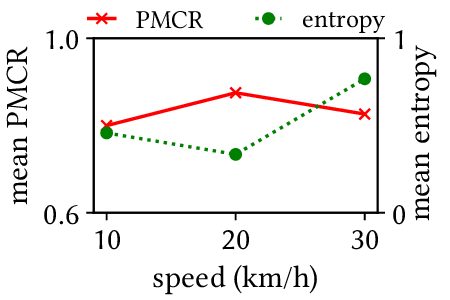}
    \vspace{-2em}
    \caption{Impact of victim vehicle's speed.}
    \label{fig:impact_of_speed}
  \end{minipage}
\end{figure}

\subsubsection{Impact of distance.}

\label{subsubsec:distance}
We use \emph{GS2} to understand the impact of sign-vehicle distance on the attack effectiveness.
We examine how the attack effectiveness metrics vary with the distance between the moving vehicle and sign. We split the road section to 22 one-meter segments, and calculate the metrics within each segment. Fig.~\ref{fig:impact_of_distance} shows results. When the camera first perceives the traffic sign, the MR can reach 77.6\% but the PMCR is low at 46.7\%. However, as the vehicle moves closer to the traffic sign, both the MR and PMCR increase. Within an distance of $25\,\text{m}$, both the MR and PMCR remain high above 97\% and 80\%. 
The degradation of attack effectiveness at farther distances are possibly due to the attenuated attack light intensity. 
Besides, the longer the distance, the smaller the $N_{sign}$, and the more vague the stripes on the sign in the FoV. This is because the time difference between the exposure of two adjacent vertical portions in a sign is smaller. 
Consequently, the light signal at each moment has more similar effects on these adjacent portions.
The performance degradation may be mitigated by increasing the intensity of the attack light (e.g., increase the LED power or use spotlight). 
Besides, perception results nearer to the traffic sign may be more significant to driving decision making, because earlier perception results may be overwritten by newer ones.

\subsubsection{Impact of movement speed.}

We use \emph{GS2} to study the impact of vehicle movement speed on the attack effectiveness. We test with speeds at around $10$, $20$, and $30\,\text{km/h}$, separately.
Fig.~\ref{fig:impact_of_speed} shows the mean PMCR and entropy vesus vehicle speed. We do not observe noticeable relationship between the attack performance and speed.

\subsubsection{Sign classes \& white/black-box attack}
\label{subsubsec:sign_class_and_wb_bb}

\begin{table}[]
  \caption{Attack effectiveness on most common traffic signs. (WB: white-box, BB: black-box)}
  \label{tab:sign_classes_and_wbbb}
  \footnotesize
  \begin{tabular}{c|cc|c|ccccc}
    \cline{1-6}
    \multirow{2}{*}{\textbf{Original}}                                             & \multicolumn{2}{c|}{\textbf{MR for GS1}}                                 & \multirow{2}{*}{\textbf{Target of GS2}} & \multicolumn{2}{c}{\textbf{PMCR for GS2}}      &  &  &  \\ \cline{2-3} \cline{5-6}
                                                                                   & \multicolumn{1}{c|}{\textbf{WB}}              & \textbf{BB}              &                                         & \multicolumn{1}{c|}{\textbf{WB}} & \textbf{BB} &  &  &  \\ \cline{1-6}
    \multirow{6}{*}{Stop}                                                          & \multicolumn{1}{c|}{\multirow{6}{*}{89.8\%}}  & \multirow{6}{*}{81.9\%}  & Speed limit 20km/h                      & \multicolumn{1}{c|}{70.6\%}      & 71.7\%      &  &  &  \\
                                                                                   & \multicolumn{1}{c|}{}                         &                          & Speed limit 30km/h                      & \multicolumn{1}{c|}{99.1\%}      & 99.9\%      &  &  &  \\
                                                                                   & \multicolumn{1}{c|}{}                         &                          & Speed limit 80km/h                      & \multicolumn{1}{c|}{100\%}       & 99.2\%      &  &  &  \\
                                                                                   & \multicolumn{1}{c|}{}                         &                          & Right-of-way                            & \multicolumn{1}{c|}{96.6\%}      & 92.8\%      &  &  &  \\
                                                                                   & \multicolumn{1}{c|}{}                         &                          & Priority road                           & \multicolumn{1}{c|}{99.3\%}      & 98.9\%      &  &  &  \\
                                                                                   & \multicolumn{1}{c|}{}                         &                          & End of no passing                       & \multicolumn{1}{c|}{72.8\%}      & 22.4\%      &  &  &  \\ \cline{1-6}
    Yield                                                                          & \multicolumn{1}{c|}{54.8\%}                   & 0\%                      & Priority road                           & \multicolumn{1}{c|}{97.3\%}      & 10.1\%      &  &  &  \\ \cline{1-6}
    \multirow{9}{*}{\begin{tabular}[c]{@{}c@{}}Speed limit\\ 30km/h\end{tabular}}  & \multicolumn{1}{c|}{\multirow{9}{*}{92.3\%}}  & \multirow{9}{*}{73.3\%}  & Speed limit 50km/h                      & \multicolumn{1}{c|}{96.0\%}      & 94.1\%      &  &  &  \\
                                                                                   & \multicolumn{1}{c|}{}                         &                          & Speed limit 60km/h                      & \multicolumn{1}{c|}{84.0\%}      & 25.9\%      &  &  &  \\
                                                                                   & \multicolumn{1}{c|}{}                         &                          & Speed limit 80km/h                      & \multicolumn{1}{c|}{100\%}       & 99.9\%      &  &  &  \\
                                                                                   & \multicolumn{1}{c|}{}                         &                          & End of Speed limit 80km/h               & \multicolumn{1}{c|}{97.2\%}      & 97.5\%      &  &  &  \\
                                                                                   & \multicolumn{1}{c|}{}                         &                          & Right-of-way                            & \multicolumn{1}{c|}{88.3\%}      & 20.2\%      &  &  &  \\
                                                                                   & \multicolumn{1}{c|}{}                         &                          & \textgreater{}3.5 tons prohibited       & \multicolumn{1}{c|}{62.1\%}      & 64.3\%      &  &  &  \\
                                                                                   & \multicolumn{1}{c|}{}                         &                          & Children crossing                       & \multicolumn{1}{c|}{97.4\%}      & 92.3\%      &  &  &  \\
                                                                                   & \multicolumn{1}{c|}{}                         &                          & End speed \& passing limits             & \multicolumn{1}{c|}{96.8\%}      & 76.4\%      &  &  &  \\
                                                                                   & \multicolumn{1}{c|}{}                         &                          & Keep right                              & \multicolumn{1}{c|}{98.1\%}      & 99.6\%      &  &  &  \\ \cline{1-6}
    \multirow{14}{*}{\begin{tabular}[c]{@{}c@{}}Speed limit\\ 80km/h\end{tabular}} & \multicolumn{1}{c|}{\multirow{14}{*}{75.0\%}} & \multirow{14}{*}{70.8\%} & Speed limit 20km/h                      & \multicolumn{1}{c|}{98.6\%}      & 90.7\%      &  &  &  \\
                                                                                   & \multicolumn{1}{c|}{}                         &                          & Speed limit 30km/h                      & \multicolumn{1}{c|}{99.9\%}      & 100\%       &  &  &  \\
                                                                                   & \multicolumn{1}{c|}{}                         &                          & Speed limit 50km/h                      & \multicolumn{1}{c|}{96.9\%}      & 99.6\%      &  &  &  \\
                                                                                   & \multicolumn{1}{c|}{}                         &                          & Speed limit 60km/h                      & \multicolumn{1}{c|}{91.2\%}      & 70.9\%      &  &  &  \\
                                                                                   & \multicolumn{1}{c|}{}                         &                          & End of speed limit 80km/h               & \multicolumn{1}{c|}{90.6\%}      & 90.3\%      &  &  &  \\
                                                                                   & \multicolumn{1}{c|}{}                         &                          & Yield                                   & \multicolumn{1}{c|}{86.2\%}      & 6.2\%       &  &  &  \\
                                                                                   & \multicolumn{1}{c|}{}                         &                          & Stop                                    & \multicolumn{1}{c|}{99.6\%}      & 10.9\%      &  &  &  \\
                                                                                   & \multicolumn{1}{c|}{}                         &                          & No vehicles                             & \multicolumn{1}{c|}{74.6\%}      & 0\%         &  &  &  \\
                                                                                   & \multicolumn{1}{c|}{}                         &                          & Slippery road                           & \multicolumn{1}{c|}{72.7\%}      & 15.4\%      &  &  &  \\
                                                                                   & \multicolumn{1}{c|}{}                         &                          & Road narrows on the right               & \multicolumn{1}{c|}{89.9\%}      & 60.4\%      &  &  &  \\
                                                                                   & \multicolumn{1}{c|}{}                         &                          & Children crossing                       & \multicolumn{1}{c|}{94.9\%}      & 100\%       &  &  &  \\
                                                                                   & \multicolumn{1}{c|}{}                         &                          & Bicycles crossing                       & \multicolumn{1}{c|}{92.5\%}      & 9.6\%       &  &  &  \\
                                                                                   & \multicolumn{1}{c|}{}                         &                          & Keep right                              & \multicolumn{1}{c|}{95.5\%}      & 98.0\%      &  &  &  \\
                                                                                   & \multicolumn{1}{c|}{}                         &                          & No passing                              & \multicolumn{1}{c|}{61.9\%}      & 9.1\%       &  &  &  \\ \cline{1-6}
    \end{tabular}
    \end{table}

We evaluate the feasibility of \emph{GS} against different groundtruth and targeted classes in a stationary setting at a sign-camera distance of $16\,\text{m}$.
We select the most common signs, including ``stop'', ``yield'' and ``speed limit'' \cite{mostcommontrafficsigns}. For ``speed limit'', we select ``speed limit 30km/h'' and ``80km/h'' as examples. 
Table~\ref{tab:sign_classes_and_wbbb} lists the target classes that are semantically conflicting with white-box PMCRs over 60\%.
The target classes for \emph{GS2} are not arbitrary for each original sign. This is due to the constraints of the perturbations' stripy forms. Besides, the ``yield'' sign is harder to compromise, likely due to its distinct inverted triangle shape that are different from the others.
Still, the results show that it is possible for the attacker to design specific attack scenarios (e.g., speed-up attack, sudden-braking attack, sign-ignoring attack) against the victim according to the expected attack consequence.
The attacker can determine the feasible set of target signs by training for each semantically-conflicting sign and select the applicable ones according to the expected attack scenarios.

Table~\ref{tab:sign_classes_and_wbbb} also compares the attack effectiveness obtained under the white-box and black-box settings. The training of a black-box attack is more challenging to converge to some targeted classes than white-box attack. This is because the black-box attack faces more constraints such as stripe widths and counts. However, it is still notably feasible as it achieves high attack success rates on several targeted classes.

\subsection{Evaluation on Lab Testbed}

We investigate the impacts of various factors on \emph{GS2}. In this subsection, unless otherwise specified, we plan the attack based on a camera exposure time of $1/1000\,\text{s}$ and sign-camera distance of $2\,\text{m}$ on the testbed, which is equivalent to $20\,\text{m}$ in real world.

\subsubsection{Exposure requirement.}
\begin{figure}
  \centering
  \begin{subfigure}{0.36\linewidth}
    \vspace{-0em}
    \includegraphics[width=\linewidth]{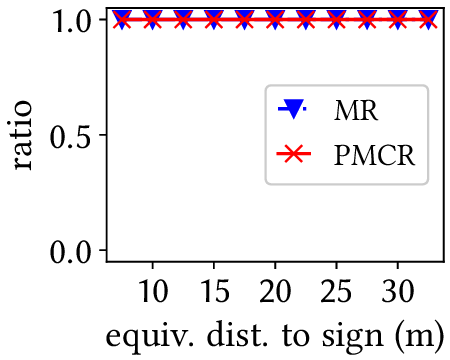}
    \caption{$t_{exp}\le1/750\,\text{s}$.}
    \label{sfig:distance_ep750plus}
  \end{subfigure}
  \begin{subfigure}{0.30\linewidth}
  \centering
    \includegraphics[width=\linewidth]{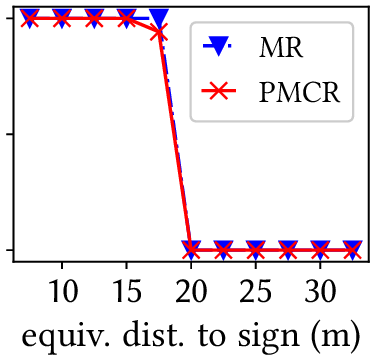}
    \caption{$t_{exp}=1/500\,\text{s}$.}
    \label{sfig:distance_ep500}
  \end{subfigure}
  \begin{subfigure}{0.30\linewidth}
  \centering
    \includegraphics[width=\linewidth]{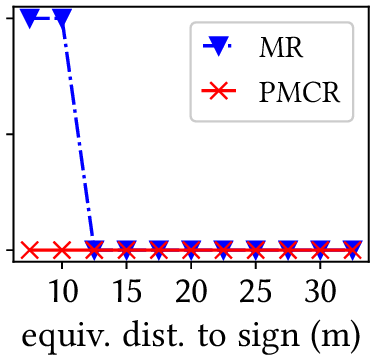}
    \caption{$t_{exp}=1/250\,\text{s}$.}
    \label{sfig:distance_ep250}
  \end{subfigure}
  \vspace{-1em}
  \caption{Effectiveness vs. sign-camera distance under different $t_{exp}$.}
  \vspace{-1em}
  \label{fig:distance_and_exposure}
\end{figure}

We use \emph{GS2} to test with exposure time $t_{exp}$ ranging from $1/2000\,\text{s}$ to $1/250\,\text{s}$ at different sign-camera distances. As shown in Fig.~\ref{sfig:distance_ep750plus}, when $t_{exp}$ is small (i.e., $\le 1/750\,\text{s}$), the PMCR is always high across a range of sign-camera distance.
When $t_{exp}=1/500\,\text{s}$ in Fig.~\ref{sfig:distance_ep500}, the PMCR is high when the equivalent sign-camera distance is shorter than $17.5\,\text{m}$. When $t_{exp}=1/250\,\text{s}$ in Fig.~\ref{sfig:distance_ep250}, the targeted attack fails at any distance as PMCR is always zero, and MR only remains high within short distances. 
This is because when $t_{exp}$ is larger, adjacent scanlines have a larger ratio of time overlaps being exposed. 
With larger $t_{exp}$ or smaller $N_{sign}$ (as discussed in \sect\ref{subsubsec:distance}), the colored stripes in a perturbation become more vague and thus less effective.
These results suggest that GS requires short $t_{exp}$ (<$1/500\,\text{s}$) at the vehicle camera to ensure successful attacks along a long distance. As autonomous vehicles are highly motion-involved, to freeze the rapid changes in the surrounding environment, a short $t_{exp}$ less than $1/500\,\text{s}$ is usually required to avoid motion blur \cite{shutterspeed2012}. Thus, the exposure requirement does not impede GS.

\subsubsection{Impact of exposure estimation bias.} 
\begin{figure}
  \centering
  \begin{minipage}[t]{0.48\linewidth}
    \centering
    \includegraphics[width=\textwidth]{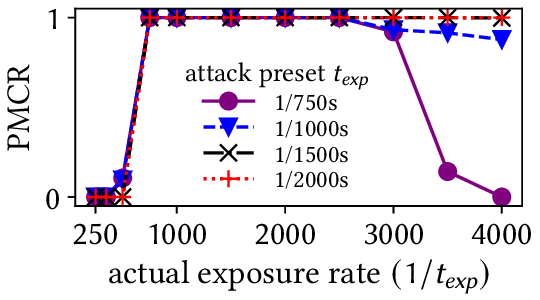}
    \vspace{-2em}
    \caption{Effectiveness with exposure bias.}
    \label{fig:exposure_bias}
  \end{minipage}
  \hfill
  \begin{minipage}[t]{0.48\linewidth}
    \centering
    \includegraphics[width=\textwidth]{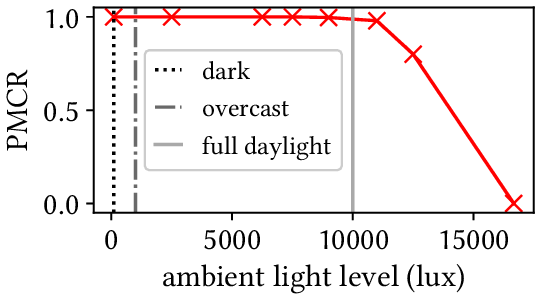}
    \vspace{-2em}
    \caption{Effectiveness vs. ambient light.}
    \label{fig:lighting_conditions}
  \end{minipage}
  \vspace{-1em}
\end{figure}

We use \emph{GS2} to study the tolerance to exposure estimation bias. We prepare the attacks for different exposure times $t_{exp}$, i.e., $1/750\,\text{s}$, $1/1000\,\text{s}$, $1/1500\,\text{s}$, and $1/2000\,\text{s}$. Then, we test them with different actual $t_{exp}$ on the victim camera.
Fig.~\ref{fig:exposure_bias} shows the PMCR under exposure estimation bias. All four attack exposure settings perform well within wide ranges of the actual exposure, showing the robust attack effectiveness against exposure bias.
The exposure bias can affect the differences between the desired and actual perturbation sharpness, size and the overall image brightness.
First, when the actual $t_{exp}$ is larger than $1/500\,\text{s}$, the attack PMCR is low due to the poor perturbation sharpness.
Second, the perturbation size defined by the duration attack window is affected by the bias in $t_{exp}$. 
When the actual $t_{exp}$ is within the working range (i.e., $<1/500\,\text{s}$), as the $t_{exp}$ is already small, the introduced size error is usually small and tolerable. 
Third, camera exposure affects the amount of input light, resulting in differences in image brightness between training data and run-time images. Large mismatches in exposure may cause large brightness difference and reduce the attack effectiveness.

\subsubsection{Impact of lighting conditions.}
As it is hard to control the ambient light outdoors, we use controllable light sources indoors to study the relationship between the attack performance and lighting conditions.
We use two studio lamps to change the ambient light level to mimic different light levels outdoors. Fig.~\ref{fig:lighting_conditions} shows the attack effectiveness under different ambient lighting conditions measured on the traffic signs with reference to outdoor conditions \cite{lightlevels}. With stronger ambient light, the attack performance decreases. This degradation occurs because the attack light is overwhelmed by the ambient light. Therefore, with brighter ambient light, the attack light needs higher power. Besides, this suggests that the attacker may need to consider the time and location when planning the attack, e.g., avoid those where direct sunlight shines on the sign (usually over 100,000 lux). 
Note that in \sect\ref{subsec:eval_outdoor}, we have demonstrated the attack effectiveness of \emph{GS} under normal daytime ambient light conditions.

\section{Possible Countermeasures}

\label{sec:countermeasures}

There are several countermeasures that may be applied to counteract the GhostStripe attack.

\textbf{Camera exposure mechanism.}
A straightforward way is to replace the widely used rolling shutter cameras by global shutter cameras.
Another countermeasure is to shuffle or randomize the sequence of scanline exposure \cite{gu2010coded,vera2022shuffled}, which spreads the attack pattern to various scanlines different from the desired perturbation. However, such countermeasures impose new requirements and extra costs on the manufacturers of autonomous vehicles and cameras, and may not be feasible for all autonomous vehicles.

\textbf{Attack-resistant perception models.}
One way to improve the robustness is adversarial training. At the training phase of the recognition models, the autonomous driving system engineers can include the labeled attack-disturbed images into the training data. This might help improve the trained model's resistance to the attack. However, this countermeasure requires significant data collection. The adversarial training may also degrade the recognition performance in the absence of attack.

\textbf{System-level redundancy.}
Multi-camera coordination may help mitigate the attack effect. Since GhostStripe is designed against a single camera, it is usually not effective against other cameras with different specifications (e.g., focal length, exposure, sensor size, altitude). However, in many autonomous vehicle solutions, there is a hierarchical camera coordination scheme. For example, the traffic light recognition in Baidu Apollo uses the output from the telephoto camera in priority, and uses the wide angle camera with shorter focal length as the backup \cite{apollotrafficlight}. In this case, the attacker can still focus on attacking the main camera.
Another possible countermeasure is to use digital maps such as High-Definition (HD) map to assist the perception of traffic sign. The autonomous vehicle can obtain the traffic signs' semantics and locations labeled in the digital map.
However, the construction, updating, and scaling of HD maps and the labeling of all the traffic signs on the map can be expensive and time consuming \cite{roadtoeverywhere}, which reduces the desirability of the map-based countermeasure. 
Moreover, maps may not cover all areas, especially in rural or remote areas, and may not adapt to changes in traffic signs due to say {\em ad hoc} construction or special events.

\section{Limitations and Discussions}

\label{sec:discussion}

\textbf{Physical access for sniffer installation.} The requirement of physical access for sniffer installation may limit GhostStripe2's opportunity.  
A determined adversary could potentially obtain the physical access by collaborating with an auto-care provider for installation. Alternatively, attackers may resort to GhostStripe1 for untargeted attacks. Exploring real-time remote sensing or eavesdropping for camera operation is an interesting future work direction.

\textbf{Attack practicability under different conditions.} Our prototype achieves similar scales as prior works \cite{ji2021poltergeist,yan2022rolling,cao2021invisible} and show high attack chances. For longer ranges and stronger ambient light conditions, the attacker may need to adopt brighter LEDs. For very high victim vehicle speed, the system latencies (e.g., from vehicle tracker and camera sniffer to the LED controller) may need to be further reduced.

\textbf{Autonomous driving system-level evaluation.} 
As the traffic sign recognition results are used by a driving agent to make decisions, it is interesting to understand whether the misled results, which may not be fully stable as shown in our evaluation, can lead to safety incidents. Using simulations is probably the only safe way to study this. However, to the best of our knowledge, publicly available driving agents only deal with traffic lights, but not traffic signs sensed at run time. Future work addressing this gap, which requires the construction of a full-fledged publicly accessible driving agent, is meaningful.

\textbf{Black-box optimization efficiency.} Our experiment reveals that while black-box attack is feasible, its BO-based low cardinality optimization falls short compared with the white-box attack. Specifically, it is more challenging to converge well for some target classes due to the constraints of stripe widths and counts. Although the attacker may prepare the attack offline with numerous queries, it is desired to obtain the attack vector towards specific target classes more effectively and efficiently. Other black-box optimization methods such as \cite{chen2017zoo, tu2019autozoom,liu2016delving,dong2019evading} may further strengthen the black-box attack.

\textbf{Other car-borne cameras.} 
In \sect\ref{subsec:att_sync}, we consider multiple commercial off-the-shelf cameras to show that the magnetic emanations from camera cables are generally indicative of the framing moments. In the real-world implementation, we only evaluate the Leopard Imaging AR023ZWDR camera because it is the default main camera used in Baidu Apollo autonomous driving system \cite{apollohardware} and the only one used for vehicles. Evaluating the proposed attack against more cameras used by various vehicles is of great interest.

\textbf{Study on human awareness.} 
While GhostStripe operates at a flickering rate invisible to human eyes, the awareness of human observers regarding the attack can be further studied. Such a study should involve human subjects to rate the suspicion levels of traffic signs under various settings, e.g., no instrumentation, truly benign illumination, malicious light flickering, and malicious stickers/paintings. 

\textbf{Single-vehicle attack.} 
GhostStripe customized the attack light signal modulation for a specific vehicle model, requiring knowledge of the victim camera specification and DNN access. It can compromise only one vehicle in the considered model approaching the traffic sign at a time, not multiple such vehicles on different lanes simultaneously.

\section{Related Work}
\label{sec:related_work}

\textbf{Physical attacks on autonomous vehicle camera perception.}
There are two classes of physical attacks, i.e., {\em object perturbation} and {\em camera perturbation}. Object perturbation attacks modify the appearance of the objects, including paper stickers and light pasted/projected onto traffic sign to mislead sign recognition \cite{eykholt2018robust,lovisotto2021slap}, painting on roadside billboard to mislead steering angle \cite{zhou2020deepbillboard}, 3D-printed object to escape detection \cite{cao2021invisible}, dirt-like patch or small marks on road surface to mislead lane detection \cite{sato2021dirty,jing2021too}, and depth-less images recognized as real objects \cite{nassi2020phantom}. All the above attacks are visible to human eyes.
Camera perturbation attacks exploit the camera hardware properties, e.g., using lasers to blind the camera \cite{petit2015remote,yan2016can}, projecting adversarial patterns into the camera lens by exploiting the lens flare/ghost effects \cite{man2020ghostimage}, using infrared light to create magenta pixels and mislead camera-based perception \cite{wang2021can}. The above camera perturbation attacks require directing the attack light into the camera lens. The related physical maneuvers are nontrivial. Differently, GhostStripe leverages the traffic sign to reflect the attack light and requires no physical maneuvers. A recent work \cite{sato2024invisible} uses invisible infrared laser to reflect projections off a portion of a traffic sign as perturbations in purple or magenta to fail traffic sign recognition. However, it is only effective for cameras without infrared filter.
The work \cite{ji2021poltergeist} uses sound wave to interfere with the image stablizer's built-in inertial sensor and trigger unwanted motion compensation. However, it focuses on disturbing the detection of on-road objects in a single frame and does not address the attack stability requirement.

\textbf{RSE applications and exploitation for attacks.}
Many visible light communication (VLC) systems are designed based on RSE \cite{danakis2012using, hu2013lightsync, hu2015colorbars,lee2015rollinglight,yang2017ceilingtalk,yang2019composite}. Specifically, the light source encodes information into controlled flickering, while the camera extracts the information from the RSE-induced stripes. Such a VLC capability can be employed in indoor localization of smartphones with LED landmarks \cite{kuo2014luxapose,rajagopal2014visual}. RSE has also been employed to watermark a physical or film scene by flickering LED or re-encoding the film video against unauthorized photographing \cite{zhu2017automating,zhang2015kaleido}.

In addition to \cite{sayles2021invisible} that is employed as a baseline attack method in this paper,
a few other works \cite{li2020light, kohler2021they,yan2022rolling} also exploit RSE to mislead computer vision. The work \cite{li2020light} shows the possibility of RSE-based backdoor attack. Specifically, during training data collection, it uses light flickering to create RSE-induced stripes as a trigger and assign an adversarial class label to the poisoning samples. During inference, the same light flickering is used as the trigger to induce the backdoored classifier to yield the adversarial class.
The works \cite{kohler2021they,yan2022rolling} particularly consider RSE-based attacks in the context of autonomous vehicles. The work \cite{kohler2021they} models the rolling shutter process by collecting RSE patterns with various parameter settings in a dark room. Certain RSE patterns overlaid on captured images can lead to miss detection of up to 75\% objects. In an autonomous vehicle simulator, the attack can introduce noticeable braking delays when there is a pedestrian or cyclist in front of the vehicle under attack. The work \cite{yan2022rolling} uses a laser to cause a monochromatic stripe that covers the traffic light to disturb the traffic light color recognition. The emission duration of the laser is controlled based on the frame time.
However, 
these two attacks \cite{kohler2021they,yan2022rolling} require 
aiming the laser at the victim vehicle's camera lens, while GhostStripe is free of this requirement.
Moreover, the above works \cite{li2020light,kohler2021they,yan2022rolling} do not consider the phase synchronization issue discussed in \sect\ref{subsec:timing-control}. Thus, they cannot control the positions of the RSE-induced stripes. Differently, GhostStripe2 applies framing sniffer to achieve phase synchronization.

\section{Conclusion}
\label{sec:conclusion}

This paper describes GhostStripe, an attack system that exploits the CMOS camera's RSE to generate adversarial stripes to mislead the traffic sign recognition of autonomous vehicles. To achieve a stable attack, GhostStripe controls the timing of the LED's modulated light emission to adapt to the camera's operations and the victim vehicle's movement. In our experiments, GhostStripe can consistently spoof the traffic sign recognition to produce a semantic-conflicting result on consecutive frames. 
This paper also discusses possible countermeasures.

\begin{acks}
  This research/project is supported by the National Research Foundation Singapore and DSO National Laboratories under the AI Singapore Programme (AISG Award No: AISG2-GC-2023-006). We thank Junming Zeng for helping customize the LED drivers, and Changhao Tian for driving the car in the real-world experiments.
\end{acks}

\bibliographystyle{ACM-Reference-Format}
\bibliography{refs}

\end{document}